\documentclass[letterpaper,twocolumn,10pt]{article}
\usepackage{usenix2019}

\usepackage{pgf-pie}
\usepackage{algorithmic}
\usepackage[ruled,vlined]{algorithm2e}
\usepackage{url}
\usepackage{multirow}
\usepackage{amsmath}
\usepackage{dirtytalk}
\usepackage{ctable}
\usepackage{threeparttable}

\usepackage{subcaption}
\usepackage{listings}
\usepackage{multicol}
\usepackage{paracol}
\usepackage{xcolor}
\usepackage{amsfonts}
\usepackage{enumitem}
\usepackage{cite}
\usepackage{graphicx}
\usepackage{array}
\usepackage{caption}
\usepackage{fontawesome} 
\usepackage{framed}

\usepackage{float}
\usepackage{tabularray}
\usepackage[most]{tcolorbox}

\tcbset{
  compactbox/.style={
    colback=gray!10,    
    colframe=black,     
    boxrule=0.3mm,      
    arc=2mm,            
    left=1mm,           
    right=1mm,          
    top=1mm,            
    bottom=0.5mm,         
    boxsep=0mm,         
    before skip=5pt,    
    after skip=2pt,     
    breakable,
  }
}

\graphicspath{{./figures/}}

\usepackage{xcolor}
\newcommand*\rot{\rotatebox{90}}
\newcommand\like[1]{\begin{picture}(1,1)
\ifnum0=#1\put(.5,.35){\circle{0.8}}\else
\ifnum10=#1\put(.5,.35){\circle*{0.8}}\else
\put(.5,.35){\circle{0.8}}\put(.5,.35){\circle*{.#1}}
\fi\fi\end{picture}}

\newcommand\bsub[1]{\vspace{0pt}\noindent\textbf{#1}}

\begin{document}

\title{Mind the Gap: Revealing Security Barriers through Situational Awareness of Small and Medium Business Key Decision-Makers}

\author{
    {\rm Yuanhaur Chang$^{\dag}$, Oren Heller$^{\dag}$, Yaniv Shlomo$^{\dag}$} \\
    {\rm Iddo Bar-Noy$^{\S}$, Ella Bokobza$^{\dag}$, Michal Grinstein-Weiss$^{\dag}$, Ning Zhang$^{\dag}$}\\
    $^{\dag}$ Washington University in St. Louis, MO, USA \\
    $^{\S}$ Israel National Cyber Directorate
}

\maketitle
\begin{abstract}
Key decision-makers in small and medium businesses (SMBs) often lack the awareness and knowledge to implement cybersecurity measures effectively. To gain a deeper understanding of how SMB executives navigate cybersecurity decision-making, we deployed a mixed-method approach, conducting semi-structured interviews (n=21) and online surveys (n=322) with SMB key decision-makers. Using thematic analysis, we revealed SMB decision-makers' perceived risks in terms of the digital assets they valued, and found reasons for their choice of defense measures and factors impacting security perception. We employed the situational awareness model to characterize decision-makers based on cybersecurity awareness, identifying those who have comparatively low awareness in the fight against adversaries. We further explored the relationship between awareness and business attributes, and constructed a holistic structural equation model to understand how awareness can be improved. Finally, we proposed interventions to help SMBs overcome potential challenges.
\end{abstract}

\thispagestyle{plain}
\pagestyle{plain}

\section{Introduction}
\label{sec:intro}
Small and medium businesses (SMBs) contribute significantly to the global economy. According to the World Bank, SMBs represent 90\% of businesses and over 50\% of employment worldwide~\cite{world_bank}. However, cybersecurity poses significant challenges for SMBs, which often lack the resources and expertise to combat sophisticated cyber threats. With limited budgets and no dedicated IT security teams, these businesses struggle to keep up with evolving security protocols, increasing their vulnerability to data breaches and financial losses. This not only impacts their own operations but also poses risks to larger organizations they may be connected with. In Israel, SMBs are responsible for 39\% of national employment~\cite{oecd}, though they are mostly unprepared against cyber-crimes. In 2023, 33,000 SMBs have fallen victim to cyber incidents, with 21\% of those suffering major or irrecoverable damage~\cite{SMB_2023}. 

\bsub{Gap in Existing Work and Our Focus.}
Recognizing the importance of the topic, there has been a significant amount of attention to understand the security practice of SMBs~\cite{chidukwani2022survey, alahmari2020cybersecurity, chen2016cyber, de2023no, huaman2021large}. However, the focus has been either on understanding cybersecurity policy and implementations from both employees' perspectives~\cite{stegman2022my, kekulluouglu2023we} and the IT counselors' perspectives~\cite{wolf2021security}, but not from the key decision-makers themselves. While prior studies provide important information on the current state-of-the-art cybersecurity practices in SMBs, there are still gaps in translating from the measurement of security hygiene to understanding the key obstacles for maximizing the appropriate cybersecurity protection for business operations. Therefore, an in-depth understanding of decision-makers' perceptions and decisions in the context of different business characteristics is essential, particularly for implementing effective interventions to fundamentally shift the posture of SMB cybersecurity at a societal scale.  

\bsub{Key Research Questions.}
To facilitate the development of interventions and motivate key decision-makers to adopt a security-aware mindset, we aim to develop an in-depth understanding of how decision-makers make cybersecurity decisions. We constructed four research questions that drove the measurement process and the corresponding analysis:

\begin{itemize}[topsep=1pt]
    \setlength\itemsep{0em}
    \item \textbf{RQ1:} What are key decision-makers' perceived cyber threats and risks for SMBs? 
    \item \textbf{RQ2:} How do key decision-makers perceive cyber defenses and their impact on company operations?
    \item \textbf{RQ3:} What factors influence key decision-makers' security perception?
    \item \textbf{RQ4:} What perceived roadblocks hinder better security?
\end{itemize}

\bsub{Contribution 1: Specific assets, protections, and factors of influence from semi-structured interviews.}
Through our semi-structured interviews with 21 key decision makers, we identified the actions implemented and the potential challenges faced by SMBs. We inductively coded these responses, reporting themes and factors they kept in mind when directing cybersecurity implementations. The findings also served as a foundation for the quantitative study.

\bsub{Contribution 2: Correlation between business attributes and situational awareness.}
Building on the understanding from the interviews, we used quantitative analysis to further verify how the identified elements and challenges can affect situational awareness.
We recruited 322 decision-makers to understand how they perceive cyber threats in the real world. We closely examined whether and how decision-makers' perceptual awareness of cybersecurity issues is correlated with the characteristics of their businesses. We predicted the awareness issues that a business with certain characteristics would likely face, and characterized decision-makers according to their situational awareness.

\bsub{Contribution 3: Holistic SEM modeling SMB decision-making.}
We constructed a structural equation model (SEM)~\cite{hoyle1995structural} that helps to visualize the causality between factors impacting decision-maker's security mindset, drawing forth a connection between reasons and eventual cybersecurity awareness. This mapping can help future researchers develop effective interventions that tackle the obstacles faced by SMBs, enable informed decision-making, and facilitate usable business management.

\bsub{Contribution 4: Root causes and roadblocks towards secure SMB.} 
Reflecting on the semi-structured interviews as well as the findings from the survey studies and the SEM, we identified the potential roadblocks that prevent key decision-makers from reaching comprehensive situational awareness. We discussed several interventions that may be deployed to address these roadblocks, and hope to mitigate the perceptual biases and increase awareness of SMB key decision-makers.

\section{Background and Related Work}
\label{sec:relatedwork}

\bsub{Situational Awareness (SA).}
Understanding the factors affecting SMB decisions can be invaluable for creating incentives that reinforce secure behavior while dismissing misconceptions regarding cybersecurity. To this end, Renaud et al.~\cite{renaud2021cyber} extended Endsley's theory of SA~\cite{endsley1995toward} to build a framework in the cybersecurity domain. In our work, we thoroughly examined SMB key decision-makers' security awareness, delving deeper into understanding the “what” and the “why” that causes low awareness while also drawing correlations between awareness and eventual cybersecurity installment. To the best of our knowledge, our work is the first to systematically study the relationship between perceptual beliefs and business actions of SMB key decision-makers.

\begin{figure}[t]
    \centering
    \includegraphics[width=\columnwidth]{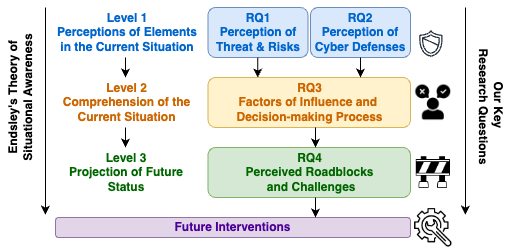}
    \caption{SMB cybersecurity with Endsley's theory of SA.}
    \label{fig:overview}
    \vspace{-15pt}
\end{figure}

Endsley’s SA theory, widely used to model human decision-making in critical situations~\cite{vieweg2010microblogging, endsley2000theoretical}, suits this study because cybersecurity threat operations require SMB decision-makers to contextualize threats/vulnerabilities according to current situations to actively defend their business. Previous studies have used SA models to analyze security perception and propose solutions in the context of eHealth~\cite{bellekens2016pervasive}, network security~\cite{martin2020contemporary, zhou2013netsecradar}, scamming scenarios~\cite{jaeger2021eyes}, or mixed reality systems~\cite{cheng2023exploring}, though none focused on SMBs. As shown in Figure~\ref{fig:overview}, our RQs can be mapped to the three levels of SA.
The first level of Endsley's theory is \textit{perception of elements in the current situation}, which from the perspective of cybersecurity, involves both the threat model and security mechanisms. Therefore, RQ1 and RQ2 are designed to gain a better understanding of business decision-makers' perceptions of these two important elements. 
Building on the perception of elements, the second level is the \textit{comprehension of the current situation}, which puts together the perceptions of all the elements in the context of SMBs. Therefore, RQ3 targets the decision-making process, which requires comprehension and composition of all the key elements in the current cybersecurity situation. 
The last level focuses on the \textit{projection of future status}. Understanding the gap is important, but coming up with actionable steps to improve the status quo is the ultimate goal. RQ4 focuses on understanding the challenges and plausible paths forward to address SMB's security challenges.

\bsub{SMB Cybersecurity.}
Prior work often stresses the stringent need for cybersecurity research in SMBs, pointing out that it is imperative for SMBs to have the ability to detect, respond, and recover from cyberattacks~\cite{chidukwani2022survey, alahmari2020cybersecurity}. 
For instance, Chen et al.~\cite {chen2016cyber} discussed the current state of SMBs and how they interact with emerging cyber threats, as well as various regulations currently in place and the changes necessary to ensure compliance among businesses. 
De Smale et al.~\cite{de2023no} studied how organizations 
condensed and filtered known vulnerability information. Unfortunately, the result was that no organization tried to acquire a comprehensive view of published vulnerabilities, but relied on a single source. 
While these works offered a view of the security practices of SMBs, they did not consider the diverse intrinsic characteristics of the businesses under study. Meanwhile, our work bridged the gap by investigating the panoply of SMBs in the full spectrum of technology exposure and business attributes.

\begin{table*}[t]
\centering
\caption{Demographics of SMB Key Decision-Makers and Business Characteristics}
\label{tab:interviewDemographics}
\renewcommand{\arraystretch}{0.8}
\resizebox{\textwidth}{!}{%
\begin{tabular}{clcl|ccc|cccc|lccc}
\specialrule{.2em}{.1em}{.1em}
\multicolumn{1}{l}{} & \multicolumn{3}{c}{\textbf{Business Characteristics}} & \multicolumn{3}{c}{\textbf{Interviewee Info.}} & \multicolumn{4}{c}{\textbf{Operational Aspect}} & \multicolumn{4}{c}{\textbf{Technological Aspect}} \\
\textbf{\#} & \textbf{Economic Sector} & \textbf{*Size} & \textbf{Digital Assets} & \textbf{\rot{Yrs. of Exp.}} & \textbf{\rot{Gender}} & \textbf{\rot{IT Exp.}} & \textbf{\rot{\parbox{2cm}{Activity \\Abroad}}} & \textbf{\rot{\parbox{2cm}{Regulation \\Requirements}}} & \textbf{\rot{\parbox{2cm}{Outsourced \\Sec. Consult}}} & \textbf{\rot{WFH}} & \textbf{Website} & \textbf{\rot{Ecommerce}} & \textbf{\rot{Cloud Storage}} & \textbf{\rot{CRM}} \\ \specialrule{.2em}{.1em}{.1em}
P1 & Accommodation and food services & M & Employee Data & 6 & M &  &  &  & \checkmark &  & Informational &  & \checkmark & \checkmark \\
P2 & Manufacturing & S & Operational Data & 22 & F &  &  &  & \checkmark &  & Informational &  &  & \checkmark \\
P3 & Administrative and support service activities & S & Customer Data / Operational Data & 6 & M &  &  &  & \checkmark &  & - &  & \checkmark &  \\
P4 & Financial and insurance activities & S & Customer Data / Operational Data & - & F &  &  &  & \checkmark & \checkmark & - &  & \checkmark & \checkmark \\
P5 & Financial and insurance activities & S & Customer Data & 9 & M &  &  &  & \checkmark &  & - &  & \checkmark & \checkmark \\
P6 & Construction & M & Intellectual Property / Operational Data & - & F &  &  &  & \checkmark &  & - &  & \checkmark &  \\
P7 & Information and communication & S & Intellectual Property / Customer Data & - & M &  & \checkmark &  &  & \checkmark & Informational / Online Service &  & \checkmark & \checkmark \\
P8 & Manufacturing & S & Operational Data & 29 & M &  & \checkmark &  & \checkmark &  & Informational / Online Service & \checkmark &  &  \\
P9 & Information and communication & S & Intellectual Property & - & M & \checkmark & \checkmark &  &  & \checkmark & Informational &  & \checkmark &  \\
P10 & Wholesale and retail trade & S & Operational Data & 20 & F &  &  &  & \checkmark & \checkmark & Online Services & \checkmark & \checkmark &  \\
P11 & Information and communication & S & Customer Data & - & F &  & \checkmark & \checkmark &  & \checkmark & Online Services &  & \checkmark &  \\
P12 & Professional, scientific and technical & S & Operational Data & 16 & M &  & \checkmark &  & \checkmark & \checkmark & Online Services &  & \checkmark &  \\
P13 & Accommodation and food services & S & Customer Data & 10 & M &  &  & \checkmark & \checkmark &  & Online Reservations & \checkmark & \checkmark &  \\
P14 & Professional, scientific and technical & M & Employee Data / Operational Data & 5 & F & \checkmark &  &  &  & \checkmark & Informational &  & \checkmark &  \\
P15 & Accommodation and food services & S & Customers data & 25 & F &  &  &  & \checkmark & \checkmark & Online Reservations & \checkmark & \checkmark &  \\
P16 & Professional, scientific and technical & S & Customer Data & 20 & M & \checkmark &  &  &  & \checkmark & - &  & \checkmark & \checkmark \\
P17 & Information and communication & S & Customer Data & 18 & M & \checkmark &  & \checkmark &  & \checkmark & Informational &  & \checkmark & \checkmark \\
P18 & Manufacturing & M & Operational Data & 15 & M & \checkmark & \checkmark &  &  & \checkmark & Informational &  & \checkmark & \checkmark \\
P19 & Professional, scientific and technical & S & Intellectual Property / Operational Data & 4 & M & \checkmark & \checkmark & \checkmark & \checkmark & \checkmark & Informational &  & \checkmark & \checkmark \\
P20 & Information and communication & M & Intellectual Property / Operational Data & 2 & M & \checkmark &  & \checkmark & \checkmark & \checkmark & Online Services &  & \checkmark & \checkmark \\
P21 & Manufacturing & M & Operational Data & 23 & M & \multicolumn{1}{l|}{} &  &  & \checkmark &  & Commercial & \checkmark &  & \checkmark \\ \specialrule{.2em}{.1em}{.1em}
\end{tabular}%
}
    \begin{tablenotes}
    \scriptsize
    \item *Size: Small (S) - 6 to 50 employees, Medium (M) - 51 to 100 employees; WFH: Work from home
    \end{tablenotes}
    \vspace{-15pt}
\end{table*}

One line of existing work studied SMB cybersecurity through the perspective of a single stakeholder~\cite{kekulluouglu2023we,wolf2021security,stegman2022my,watad2018security,alahmari2021investigating}. 
For example, Wolf et al.~\cite{wolf2021security} uncovered security obstacles from the perspective of Chief Information Security Officers (CISOs), regarding them as third-party observers of the actions of SMBs. 
On the other hand, Stegman et al.~\cite{stegman2022my} surveyed employees' concerns over ambiguous data collection through enterprise security software. 
Recognizing that SMB executives often juggle multiple issues that could affect the fate of the company~\cite{osborn2018risk}, an in-depth understanding of SMB cybersecurity efforts from the key decision-maker's perspectives is essential. However, findings from previous research lack context for cybersecurity decision-making and their impact on decision-makers' responses to cyberthreats is unknown~\cite{watad2018security,heidt2019investigating,alahmari2021investigating,huaman2021large}. To this end, our work explored factors and challenges influencing decision-makers' perceptions of their cybersecurity status. We addressed the lack of adequate sample size in prior work by conducting large-scale studies with a diverse set of SMBs and key decision-makers.
\section{Interview Study}
\label{sec:interview}

To understand how decisions are made in SMBs and to obtain a framework for the main survey development, we conducted an interview study exploring how SMB executives navigate cybersecurity decision-making. Our study was formally approved by the Institutional Review Board (IRB).

\bsub{Recruitment Method.}
Our study was conducted in Israel. To capture a representative sample of key decision-makers whose business belongs to different economic sectors as defined by ISIC Rev 4 classification~\cite{united_nations} of the United Nations, we recruited through commercial records while meeting the definition of SMBs~\footnote{Businesses with 6-100 employees and 10M-100M NIS annual revenue.}. We define key decision-makers as people who hold the mandate for cybersecurity policies in the business, including owners, CEOs, CTOs, Vice Presidents, and managers who report directly to the CEO or the owner of the business. We carefully selected the key decision-makers and businesses, ensuring that our sample was diverse and representative in terms of SMB characteristics. Businesses that are not privately owned were excluded. In the end, 21 key decision-makers were recruited. Participants were happy to volunteer and were not compensated. Demographics of decision-makers and their businesses are shown in Table~\ref{tab:interviewDemographics}.

\bsub{Interview Process.}
We followed a semi-structured interview protocol for the study, allowing the interviewer and the interviewees to raise and explore new issues when possible. After obtaining the participant's informed consent, the interviewer would ask questions using the interview guide in Appendix~\ref{app:interviewGuide}.

\bsub{Limitations.}
Participants may be subjected to self-selection bias. There may also be potential self-report bias, where discrepancies may exist between the decision-makers' perception and the actual situation. Social desirability bias may also cause decision-makers to report better security practices than they actually have to make their business look better. 

\bsub{Ethical Considerations.}
Before the interview, participants were given the research description and signed a consent form. All interviews were conducted in Hebrew and recorded. The recordings were then transcribed and translated by the research team for analytical purposes. Confidentiality and anonymity were given careful attention, and each participant in the study received a unique research ID number. We refrained from including identifiable information in our results.

\bsub{Thematic Coding.}
We deployed thematic analysis~\cite{braun2006using} to identify themes that help answer our research questions. Two coders independently went over the transcripts, noting and refining the initial set of themes and codes. The themes and codes were then discussed iteratively and the differences were resolved until all coders reached an agreement on the final codebook, which is presented in Appendix~\ref{app:codebook} and used to code all responses. We use this categorization in the second part of the study for survey development and quantitative analysis.
\section{Interview Results}
\label{sec:qual}

In this section, we summarize key decision-makers' perceived risks based on digital assets (\S~\ref{sub:risk}) and defense deployment (\S~\ref{sub:defense}), which are important elements to discuss for level 1 situational awareness. We summarize key decision-makers reasons for the chosen defenses, as well as the factors influencing their security perceptions (\S~\ref{sub:sources}, \S~\ref{sub:factors}). 
Through these, we identify various root causes of insecure SMBs.

\subsection{What digital assets are SMB key decision-makers concerned about?}
\label{sub:risk}

Security is ultimately a process of risk minimization, since there is no perfectly secure system. As a result, the first step towards helping SMB decision-makers achieve better security is to understand what are the essential assets they deem valuable and wish to protect. This gives us a perspective of how decision-makers perceive their current cybersecurity threats.

\bsub{Customer data for secure services.} The most prevalent information SMB executives deem as important digital assets are customer profiles and data. For individual customers, SMBs may need to securely preserve \textit{"delivery certificates and the contract of the services (P3)"} up to a certain time. In addition, for SMBs working in the healthcare sector, the security of personal health information is of great concern. P11 noted, \textit{"Theoretically, someone could break into our system and change the instructions for the patient and cause the patient to be treated incorrectly."} For customers who are companies, sensitive financial information may leak out due to malpractice or attacks. For instance, P4 expressed concern in handling customer bank credentials for tax purposes, \textit{"I have 300 clients, most of them companies. I need to log into the bank account. I received a password, and some of them gave access not only to viewing but also to making transactions. Even if not maliciously something can happen."} 

\bsub{Employee data for efficient management.} P1, who owns a restaurant, indicated that he heavily relies on apps to manage his restaurant. The apps allow him to efficiently manage employees, shifts, and salaries, helping him minimize managerial costs. \textit{"For me, the data is a major asset. In the first years before I had this data gathered things were more challenging."}

\bsub{Operational data for service availability and safety.} Some decision-makers stated that assets essential to company service should be protected since the lack or leakage of those can cause major operational issues. Many participants mentioned having a website to promote their business or as a means of communication with customers. The availability of websites is particularly vital for SMBs who utilize them as major channels for customer interaction. For P12, who runs a survey company, \textit{"A server crash in our company in the past silenced my activity for a few hours. In our world, this is critical because usually within 24 hours the survey needs to be closed and the information received."} 
Meanwhile, in a factory setting, P18 is worried about the access control of their operational technology. \textit{"There are quite a few things here, from sophisticated machines to raw materials. It definitely needs to be protected and if someone gets into [the system] they can activate a lot of things."}

\bsub{Intellectual properties for business competitiveness.} Besides the digital assets mentioned above, SMBs often have intellectual properties or business secrets that they need to protect. P6, who is the owner of a construction company, worried that their engineering plans will be stolen. In addition, owners who work in the information and communication sector expressed more concerns over the algorithms in their software development projects than customer information, stating \textit{"Mainly the code [should be protected] because we don't have customer information that could expose us to lawsuits. The fact that you work with a client is no secret." (P9)}

\begin{tcolorbox}[compactbox]
        \textit{\textbf{Discussion:}} These four major categories mentioned in participants' responses point out the different concerns decision-makers have according to different types of assets. These assets are seen to be directly related to the working and availability of company services, and if compromised, may lead to significant financial loss. A key observation is that a business with more diverse digital assets may face a larger threat surface. In other words, the \textit{technological intensity} of a business can affect decision-makers' perception of threats and risks.
\end{tcolorbox}

\subsection{What defensive measures do SMB key decision-makers choose (not) to deploy?}
\label{sub:defense}

Also related to decision-makers' level 1 situational awareness is the perception of security mechanisms and the reasons behind deployment decisions. Understanding the thought process behind defense decisions can help identify where potential misconceptions may need to be corrected, as well as where knowledge may be lacking to improve security defenses.

\bsub{Backups are important for operation.} When asked about how the company protects its digital assets, almost all the participants reflected on either having local and remote backups or hosting all of their services on the cloud. P3 said, \textit{"[Everything] is saved on local drives and on the cloud. Everything is also printed and saved in binders."} However, other than stating this defensive measure, we observed that most decision-makers do not care to understand the details of the operation. In general, participants tend to have a false sense of security about hosting their service on the cloud, believing that whatever is on the cloud is backed up and secure. P14 shared his strategic consulting experience and concurred, \textit{"Even when it is possible to negotiate terms of backup from the providers, customers are not aware of their options."}

\bsub{Divided opinion on employee training.} Some SMB executives require their employees to receive training or follow certain rules while handling business operations. For instance, P18's company conducted "mock attacks" to familiarize employees with phishing scams. \textit{"Lectures are quite boring, in my opinion, you don't take anything from it, at best you remember some nice gimmick. That's why what we do is send scams from an external email and then check who fails."} In addition, P14 spends a great effort raising security awareness among the employees, sending out monthly newsletters to employees to update them on recent incidents and requiring employees to provide comments and feedback.

On the other hand, some business owners adopted a more fatalistic viewpoint and believed there was no value in implementing employee training, as it was too difficult even to identify the source of the issue. Owners who made this decision are eager to "get back to normal". P5 reasoned, \textit{"Security is always delegated to professionals. We didn't see any point [to do training] because we can't help and the attack already happened. We just wanted to return the office to function."}

\bsub{Minimal effort on firewall, antivirus software, and guideline implementation.}
Only decision-makers who are more tech-savvy or have a higher security awareness would allocate budgets annually for cyber defenses, such as setting up firewalls and renewing antivirus software licenses, while following security standards if the nature of their company demands so. P20, who runs a software company, mentioned having developed incident response plans with scenarios that allow all employees and management to understand what to do if the company is being attacked. Moreover, P16 shared his opinion as to why some SMBs neglect to renew their antivirus licenses, \textit{"They are not stingy. They simply save every shekel because small businesses in Israel are suffocating from the economic burden. They want to see the security people work because otherwise, they don’t feel comfortable paying."}

\begin{tcolorbox}[compactbox]
    \textbf{\textit{Discussion:}} While decision-makers understand that some degree of defensive measures needs to be deployed, they tend to only do the bare minimum to save the already strained budget and time. There also exist large misconceptions about the extent or the effectiveness of the employed protections, such as cloud services are guaranteed to be safe and secure. The root cause of this oversight could be largely due to the \textit{lack of technological orientation and innovation}.
\end{tcolorbox}

\subsection{What sources of information do SMB key decision-makers rely on?}
\label{sub:sources}

Apart from digital assets and security measures, sources of decision-makers' cybersecurity knowledge can influence how they comprehend SMB's security status and make decisions, which is captured by Endsley's level 2 situational awareness. 
As seen in \S~\ref{sub:risk} and \S~\ref{sub:defense}, much of the decisions on defense and their effectiveness rely on accurate comprehension of threats by the key decision-makers.
Therefore, it is imperative to identify sources that may be unreliable.

\bsub{Expert Guidance.} Some key decision-makers seek advice from or outsource the task to dedicated agencies specializing in computer services. We observed that the frequency of interaction between SMB and the agency is surprisingly low, mostly reporting to be \textit{"once every six months (P10)"} or on-demand: \textit{"From time to time I pester them with some question at the request of a client regarding their security systems. (P12)"} 
Instead of large consulting agencies, many would choose to hire individual technicians whom someone else recommended. They expressed complete trust in the technicians, agreeing to whatever they advised. For example, \textit{"He sends me an email and I don't understand but I tell him yes. These are amounts like 30 or 50 Shekels per month. (P4)"} 

Others suggested that when the company merged with another institution, they get to know how the other party implements defensive measures. Mainly, \textit{"We have merged with a strong tax consultancy headed by the “Institute of Tax Consultants in Israel”. The senior partners in the institute accumulated lots of security know-how. We can consult on all kinds of questions such as where to improve the cyber defenses. (P5)"} 

\bsub{Structured Information Source.} A few SMB executives rely on structured sources to obtain the security knowledge necessary for company operations. When asked if there are other information sources beyond meetings with IT personnel, P21 mentioned conferences and lectures, \textit{"The Association of Manufacturers had a lecture on information security, also in business forums."}
Meanwhile, some said that they will \textit{"go over the journals that are published in this field (P14)"} or \textit{"hear about other businesses in the media} (P6)" to update themselves on the current status of their business ecosystem. 

Due to the business's specific economic sector, decision-makers may be required to become familiar with related standards such as the ISO 27001 Standard. For instance, \textit{"I adopt an ISO information security standard so that the basis of the cyber requirements are familiar to us and we try to preserve and comply with them. I also use the 9001 standard which is also a quality standard (P17)."}
However, P21 mentioned that sometimes he needed to \textit{"route between all the advice that exists in the market, which can be contradictory to one another."} He noted that \textit{"someone should make some characterizations of several levels of companies and explain what each level should do for cybersecurity."} 
Furthermore, P12 believed that having stricter regulation and enforcement could help raise awareness. \textit{"If there was an orderly definition of regulation and even tests and penalties by government bodies, then I would be more committed to it. I would have a guide that I would follow and know if I am working correctly."}

\bsub{Personal Background and Experience.} Some key decision-makers we interviewed have educational backgrounds in IT, and they mentioned using personal expertise as a source for security judgment. Interestingly, three key decision-makers attribute their IT knowledge to their time during military service. As P16 said, \textit{"All my life I studied and worked in the field of computers, not in academia, graduated from a computer unit in the army, both at the programming level and at the IT level. I learned everything from zero."}

Others said they gradually become familiar with cybersecurity through years of experience in operating the business, especially after their first encounter with cyberattacks. P8 said, \textit{"We went through a ransomware attack, the computers were locked, they asked for money, 30 bitcoins. At the time I didn't understand what Bitcoin was at all. As far as we were concerned, we understood that we had entered into a war with terrorists."} P16, who owns a company that provides IT services, also said, \textit{"I don't go to courses or further training, we learn while working, while dealing with problematic activities that have been identified with the customers."}

\begin{tcolorbox}[compactbox]
    \textbf{\textit{Discussion:}} Due to the lack of credible sources of consultation, \textit{difficulty in information navigation} can be one of the root causes for SMB's cybersecurity barrier. Surprisingly, most SMB decision-makers prefer individual technicians over consulting agencies to cut personnel expenses, and even when they have invested in large agencies, the interaction is infrequent.
    Too much information can overwhelm key decision-makers as they lack the means to effectively filter and identify useful ones. It is also noted that many choose to rely on their own experiences in cybersecurity, particularly as a victim of an incident.
\end{tcolorbox}

\subsection{What factors impact SMB key decision-makers' security decision?}
\label{sub:factors}

Finally, we look into the decision-making process after all elements are jointly considered, in which we disclose the reasons key decision-makers give for whether security measures are implemented. Understanding the rationale behind these decisions can help policy-makers and stakeholders devise suitable strategies that promote cyber defense installments.

\bsub{Whether risks are covered by another entity.}
From our interviews, we observed that executives tend to be more indifferent toward security issues when the risk can be offloaded to or mitigated by another agency or institution. While this includes hiring third-party consultants to assist the process as described in \S~\ref{sub:sources}, responsibilities in the case of an attack can also be completely shifted. P1 argued, \textit{"I don't think about cyber risks. The financial risk of payments is taken care of by the credit card company. The credit card company gives us insurance."} Also, as P5 said, \textit{"We would contact the Israeli IRS and tell them that we lost information in a ransom attack. We would continue to work and not close the business."}

\begin{table*}[h!]
\centering
\scriptsize
\caption{Demographic of Survey Participants and Businesses (N=322)}
\label{tab:surveyDemographic}
\resizebox{\textwidth}{!}{%

\begin{tabular}{|clll|cllcll|}
\hline
\multicolumn{4}{|c|}{\textbf{Business}} & \multicolumn{6}{c|}{\textbf{Decision-makers}} \\ \hline
\multirow{3}{*}{\# of Employees} & 6-10 & 26.70\% & 50\%* & \multirow{4}{*}{Position} & Owner & \multicolumn{1}{l|}{7.80\%} & \multirow{2}{*}{Gender} & Male & 54.00\% \\
 & 11-50 & 55.00\% & 44\%* &  & CEO & \multicolumn{1}{l|}{7.80\%} &  & Female & 46.00\% \\ \cline{8-10} 
 & 51-100 & 18.30\% & 6\%* &  & Vice President & \multicolumn{1}{l|}{12.70\%} & \multirow{6}{*}{\begin{tabular}[c]{@{}c@{}}Seniority \\ (years)\end{tabular}} & 1-4 & 12.70\% \\ \cline{1-4}
\multirow{6}{*}{Economic Sector} & Services & 31.40\% & 39\%* &  & Manager & \multicolumn{1}{l|}{71.70\%} &  & 5-9 & 18.60\% \\ \cline{5-7}
 & Professional services & 28.00\% & 18\%* & \multirow{5}{*}{Age} & 25-34 & \multicolumn{1}{l|}{25.20\%} &  & 10-14 & 20.20\% \\
 & Trade & 9.30\% & 27\%* &  & 35-44 & \multicolumn{1}{l|}{28.60\%} &  & 15-19 & 12.10\% \\
 & \multirow{2}{*}{\begin{tabular}[c]{@{}l@{}}Information and \\ communication\end{tabular}} & \multirow{2}{*}{18.90\%} & \multirow{2}{*}{7\%*} &  & 45-54 & \multicolumn{1}{l|}{26.70\%} &  & 20+ & 35.10\% \\
 &  &  &  &  & 55+ & \multicolumn{1}{l|}{19.30\%} &  & Refuse to answer & 1.20\% \\ \cline{8-10} 
 & Production & 12.40\% & 9\%* &  & Refuse to answer & \multicolumn{1}{l|}{0.30\%} & \multirow{6}{*}{\begin{tabular}[c]{@{}c@{}}Technology\\ Knowledge\end{tabular}} & Basic knowledge & 8.10\% \\ \cline{1-7}
\multirow{5}{*}{\begin{tabular}[c]{@{}c@{}}Annual Revenue\\ (NIS)\end{tabular}} & Up to 1 million & 9.60\% & \multicolumn{1}{c|}{-} & \multirow{5}{*}{Education} & High school diploma or less & \multicolumn{1}{l|}{25.50\%} &  & \multirow{2}{*}{\begin{tabular}[c]{@{}l@{}}Intermediate-level \\ knowledge\end{tabular}} & \multirow{2}{*}{44.70\%} \\
 & 1-5 million & 18.60\% & \multicolumn{1}{c|}{-} &  & Certificate & \multicolumn{1}{l|}{14.00\%} &  &  &  \\
 & 5-10 million & 13.00\% & \multicolumn{1}{c|}{-} &  & Bachelor's degree & \multicolumn{1}{l|}{37.00\%} &  & Advanced & 32.00\% \\
 & 10+ million & 18.00\% & \multicolumn{1}{c|}{-} &  & Master's degree or higher & \multicolumn{1}{l|}{23.00\%} &  & Professional & 13.40\% \\
 & Refuse to answer & 40.70\% & \multicolumn{1}{c|}{-} &  & Refuse to answer & \multicolumn{1}{l|}{0.60\%} &  & Refuse to answer & 1.90\% \\ \hline
\end{tabular}%
}
    \begin{tablenotes}
        \footnotesize
        \item *Real-world distribution of SMBs with the corresponding attribute
    \end{tablenotes}
    \vspace{-15pt}
\end{table*}

\bsub{Whether losing/leaking data entails inconvenience.}
When data leakage can cause inconvenience to business operations, participants are more likely to implement defensive measures. \textit{"The biggest headache is to restore documents and for that purpose, there are backups in all places so that if they take over or steal the backup there will be a backup somewhere else. (P2)"} Some would choose to focus on other parts of the business because there are no foreseeable risks. P8 added, \textit{"I know there is no complete solution and I don't want to bother with the issue either. Jams will always be produced, the information is not secret. There will be no harm."}

\bsub{Whether attacks hinder company operation.}
In addition to financial losses that may be the result of service downtime, a business' reputation can also be affected by cyberattacks, indirectly motivating decision-makers to allocate more resources for defense. P13 shared, \textit{"If a rumor gets out that we were attacked, then customers will stop believing in us and give us their details."} On the other hand, when the data is evaluated to be "non-critical", there is a significant drop in willingness to adopt security measures: \textit{"I don't see a financial risk. Regarding my operational data, I don't think they can wipe out information that is important to me. (P1)"}

\bsub{Whether other companies experienced attacks.}
While many key decision-makers failed to see the likelihood of being attacked, news of incidents from other businesses (particularly of the same niche) can remind them to implement defense for their own company. P4 viewed this as a defining moment for her to be more aware of cybersecurity, \textit{"I have clients, lawyers, who went through a cyber-attack, tried to fix the computers for 3 days and without success. In the end, they paid a ransom in Bitcoin. That day I moved to the cloud."} 

\begin{tcolorbox}[compactbox]
    \textbf{\textit{Discussion:}} These factors suggested that SMB decision-makers may have \textit{inadequate risk management} skills, as they tend to rely on other larger entities to mitigate cyber risks or believe that cyberattacks will not cause any damage to business operations. Others may perceive their digital assets as unimportant without having a proper risk evaluation, eventually neglecting to install necessary protection due to the \textit{lack of constructive decision-making}.
\end{tcolorbox}
\section{Online Survey}

To understand how business factors and root causes observed from the interview may impact SMB key decision-maker's situational awareness, and to understand in which area decision-makers may be less aware, we developed and conducted an online survey study to explore how SMB executives navigate cybersecurity decision-making. Our study was formally approved by the Institutional Review Board (IRB).

\subsection{Survey Protocol}
\label{subsec:design}

\bsub{Pilot.}
We piloted the survey with 20 SMB executives in batches, addressing feedback by removing redundant questions and clarifying question statements. The final survey instrument is included in Appendix~\ref{app:surveyinstrument}.

\bsub{Participant Recruitment.}
We recruited key decision-makers in Israel through Panel4All \cite{Panel4All}. Similar to the interview study, we excluded businesses that are not privately owned, as well as those that do not fit the definition of small and medium businesses. We surveyed only owners/CEOs/Vice Presidents/department managers who report directly to the CEO or the owner of the business. The survey took 20 minutes on average to complete and participants were compensated 10 NIS. Distribution of participant and business demographics are presented in Table~\ref{tab:surveyDemographic}.

\bsub{Data Analysis.}
We excluded responses that selected "Don't know/Refuse to response" to more than 70\% of the questions, resulting in a total of 322 responses. For the purpose of the analysis, economic sectors were categorized into five major groups. We measured a company's \textit{technological intensity} by the ownership of different types of digital assets and the digital technologies deployed. We referenced the Digital Intensity Index (DII) from Eurostat~\cite{dimodim_2022} with some modifications. Specifically, SMB received one point each time one of the following is true: 1) Company employs ICT experts, 2)  50\% of the employees use the Internet for work purposes, 3) Company has a website, 4) Company's website has advanced functions (order tracking, personalization, etc.), 5) Company purchases advanced cloud services (CRM, computing power, software, etc.), and 6) Company has online trading. We then took the average of the scores as the threshold. If a business's score is above average, it relies heavily on digital technology and is said to have high technological intensity. We describe our survey analysis methods in more detail in \S\ref{subsec:surveyanalysis}.

\bsub{Limitations.}
As with other survey studies, our sample distribution is limited by the participants we recruited, and there may also be self-reporting biases. Although our sample is not fully aligned with the real-world distribution of business sizes and economic sectors as indicated by the Israeli National Bureau of Statistics~\cite{inbs}, each business size and economic sector still has an adequate representation in our studied samples.
The alignment of business revenue between our sample and the real-world is unknown, as these are considered trade secrets and businesses often refuse to provide them. 

\bsub{Ethical Considerations.}
Participants were asked for their consent as part of the survey before starting. All responses were collected through self-report measures and anonymized, and participants were not required to disclose any information they did not want to share.

\subsection{Survey Analysis}
\label{subsec:surveyanalysis}

\subsubsection{Situational Awareness (SA)}
\label{subsubsec:awarenessmodel}

The bulk of our survey was designed with situational awareness model in mind~\cite{endsley1995toward}. We referenced the five-level framework for cyber situational awareness~\cite{renaud2021cyber}, and developed methodologies to measure and identify low awareness.
Each level of SA maps to specific questions to examine how perceptions and barriers affect SMB cybersecurity. Since survey responses are self-reported, we lack objective information on the potential damage to SMBs in the case of cyberattacks. However, our data enables us to infer this damage. For instance, the more digital assets an SMB possesses and the more sensitive the website functionalities, the higher the damage can be as a result of cyberattacks~\cite{Fortinet}. We relate SMBs' attributes and decision-makers' perceived potential damage, leveraging crowd wisdom in management~\cite{budescu2015identifying, muller2018wisdom} to identify those whose self-assessments were substantially lower than others at each level. With this population of low-SA SMBs, we performed a logistic regression to predict the probability of low SA if certain business attributes are present (Figure~\ref{fig:aw}).

\bsub{Level 1: Not being aware of the importance of cybersecurity to business continuity.}
This level characterizes a lack of basic understanding of cybersecurity matters. Executives lacking level 1 SA tend to underestimate possible damages faced by their company. 
To assess SMB key decision-makers' level 1 SA, we compared their self-assessments of their business's potential damage and the projected potential damage due to cyberattacks. If the decision-maker anticipates low damage but the business may face severe damage, then it is implied that the decision-maker exhibits low awareness. It should be noted that this projected potential damage is regardless of the precautions taken by the SMB.

\begin{figure*}[t]
    \centering
    \includegraphics[width=0.9\textwidth]{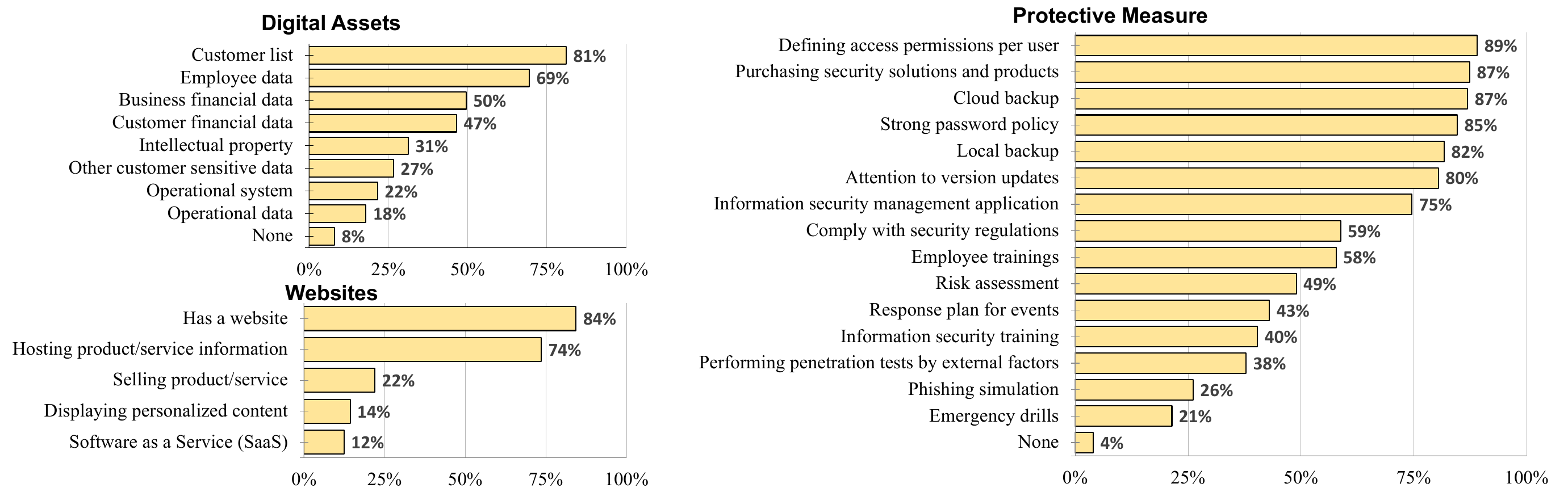}
    \caption{Percentage of SMBs owning digital assets, websites, and protective measures deployed (N = 322).}
    \label{fig:basic}
\vspace{-10pt}
\end{figure*}

To do so, we estimated this logistic regression model:
\begin{equation}
    logit(pr(Damage_i=1))=\ \beta_0+\sum\beta_j\ X_{ij}^{Level\ 1}+\epsilon_i
\end{equation}
where, $Damage_i=1$ if the answer to \textit{"In your opinion, what is the greatest possible damage that could occur in the event of the loss or theft of all the digital assets of your business?"} is either "Bankruptcy" or "Significant decrease in income/revenue" (42\%), and $Damage_i=0$ for other responses (medium/minor damage: 53\%; no damage at all: 5\%).
The variable $X_{ij}^{Level\ 1}$ includes all business attributes (number of digital assets, website functionality, number of employees, etc.) A detailed variable and coefficient table is included in Appendix~\ref{app:lv1_4_regress-coeff}.
The residual term $\epsilon_i$ represents business $i$'s deviation from the average relationship. Given a business attribute, a larger value of $\epsilon_i$ implies an overestimation of the damage and a smaller value implies an underestimation of the damage compared to other businesses. We standardized $\epsilon_i$ and used it as a measure for level 1 awareness. We defined a key decision-maker as having low level 1 awareness if its $\epsilon_i$ is of the lowest 20\%.~\footnote{Those who assessed no damage at all in case of losing all digital assets were also included as having low level 1 SA, even if they were not defined as such by the described mechanism. Those who self-assessed as anticipating severe damage were not included as low level 1 SA, even if they were defined as such by the described mechanism.}

\bsub{Level 2: Not being aware of the risk of being exposed to a cyber-attack.}
This level entails having misconceptions about the probability of being attacked. We asked, \textit{"On a scale of 1 to 10, what is the likelihood that a business like yours will be attacked in the coming year?"} This variable was standardized and used as a level 2 SA measure. Those whose self-assessment falls below the 23\% threshold (1: 12.8\% or 2: 10.6\%) were grouped as low level 2 SA.

\bsub{Level 3: Not being aware of cyber security precautions and controls.}
This level is characterized as lacking knowledge and understanding regarding the actions that need to be taken. We asked, \textit{"From a scale of 1 to 10, to what extent do you think the knowledge you have in the field of cybersecurity is sufficient?"} This variable was standardized and used as a level 3 SA measure. Those whose self-assessment falls below the 27\% threshold (1: 13.8\%; 2: 13.4\%) were grouped as having low level 3 SA.

\bsub{Level 4: Not being aware of the need to act.}
Decision-makers who have low awareness at this level may overestimate the level of protection their business has due to the misconception that the necessary defense measures have already been taken. To find out the group of decision-makers who exhibit such misconception, we first assessed to what extent are SMB precautions adequate to its needs. We estimated the following linear regression model~\footnote{Theoretically, Poisson regression would be a better model for count data; however, linear regression yielded the same results as Poisson regression in this case. For simplicity, we report results from the basic linear model.}: 
\begin{equation}
    Precautions_i=\ \beta_0+\sum\beta_j\ X_{ij}^{Level\ 4}+u_i
\end{equation}
where $Precautions_i$ is the number of protective measures of the SMB. The variable $X_{ij}^{Level\ 4}$ includes SMB attributes such as the type of digital assets, website functionalities, etc. Detailed regression variables and coefficients are included in Appendix~\ref{app:lv1_4_regress-coeff}.
The residual term $u_i$ represents business $i$'s deviation from the average number of precautions over our population sample. A large $u_i$ stands for over-cautiousness, and a low $u_i$ stands for under-cautiousness relative to other SMBs. We refer to the standardized $u_i$ as \textit{relative cautiousness}.

We next associated relative cautiousness with the subjective perception of risk. This allowed us to address decision-makers whose risk perception did not fit its cautiousness. Specifically, we wished to identify under-cautious decision-makers who believe their business is safe. We stressed a 45$^{\circ}$ line between participants' answers to the question, \textit{"On a scale of 1 to 10, what is the level of cyber protection in your business?"} (y-axis) and relative cautiousness (x-axis), both being standardized measures. Level 4 SA equals the distance from the 45$^{\circ}$ line, where decision-makers above the line show low awareness and those below show high awareness. We defined those at the lowest 20\% as having low level 4 SA.

\bsub{Level 5: Lack of resources.}
Decision-makers at this level face challenges related to the lack of resources for cybersecurity, even though they understand what needs to be done. We asked if they had encountered a lack of social influence over company personnel, or a lack of organizational resources such as the required budgets and time. Likewise, those are standardized and used as a scale for having sufficient resources. Decision-makers who reported lacking one or more resources (among budget, personnel, and time) were grouped as having low level 5 SA, which took up 25\% of the sample (lacking 1 item: 16\%; 2 items: 7\%; 3 items: 2\%).

\subsubsection{Root Causes}
In addition to evaluating the specific SA issues businesses face when implementing security measures, the survey studies whether the potential root causes identified in the interview significantly impact SA. These were asked on a 5-point scale. The responses are aggregated and the average is reported. 

\bsub{Inadequate Risk Management.}
We evaluated the adequacy of SMB risk management by asking decision-makers to rate themselves regarding the following: having a clear understanding of risks, taking active actions to reduce risks, having contingency plans, and prioritizing risk management.

\bsub{Difficulty in Information Navigation.}
We asked participants whether they felt overwhelmed or confused by the abundant cybersecurity information and whether they had difficulty staying up-to-date due to the constantly evolving threats.

\bsub{Lack of Technological Orientation.}
For technological orientation, we inquired about participants' tendency to act when implementing new technologies in the business. We also asked whether their business has explored innovative security solutions in the past three years, as well as the extent of their exposure to such solutions made by business competitors. 

\bsub{Lack of Constructive Decision-making.}
We evaluated whether business executives make constructive decisions by asking what information they base their decisions on (personal management experience, employee experiences, intuitions, external consultants, and statistical insights).

\section{Quantitative Analysis Result}

\begin{figure*}[t] 
    \centering
    \begin{subfigure}{0.47\textwidth}
        \includegraphics[width=\linewidth, trim=2 2 2 2,clip]{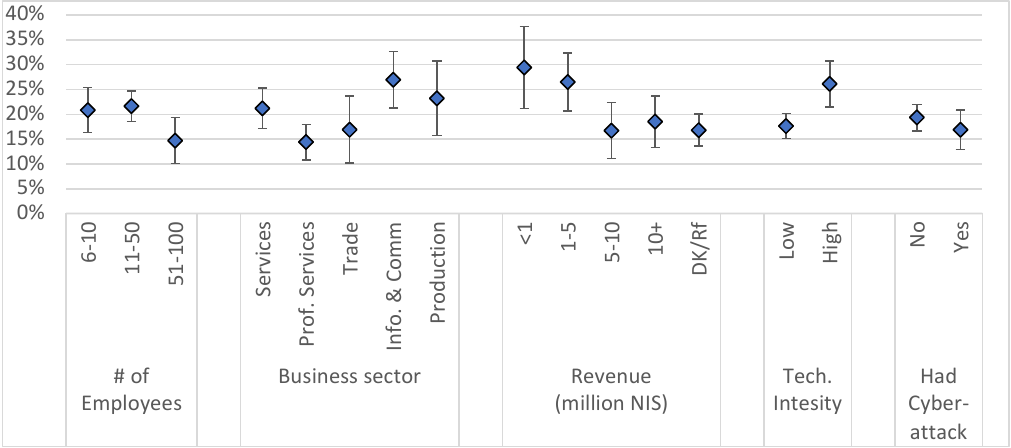}
        \caption{Low SA Level 1}
        \label{fig:aw1}
    \end{subfigure}
    \hfill 
    \begin{subfigure}{0.47\textwidth}
        \includegraphics[width=\linewidth, trim=2 2 2 2,clip]{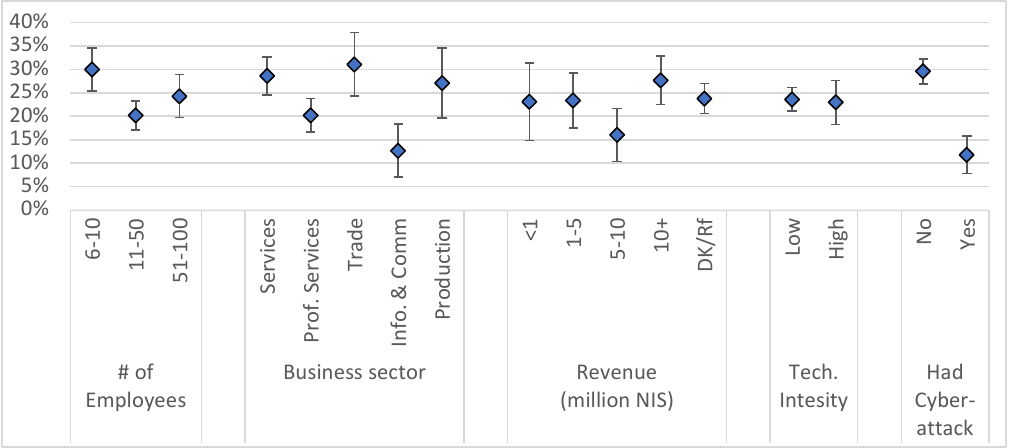}
        \caption{Low SA Level 2}
        \label{fig:aw2}
    \end{subfigure}
    \vspace{-3pt} 
    \begin{subfigure}{0.47\textwidth}
        \includegraphics[width=\linewidth, trim=2 2 2 2,clip]{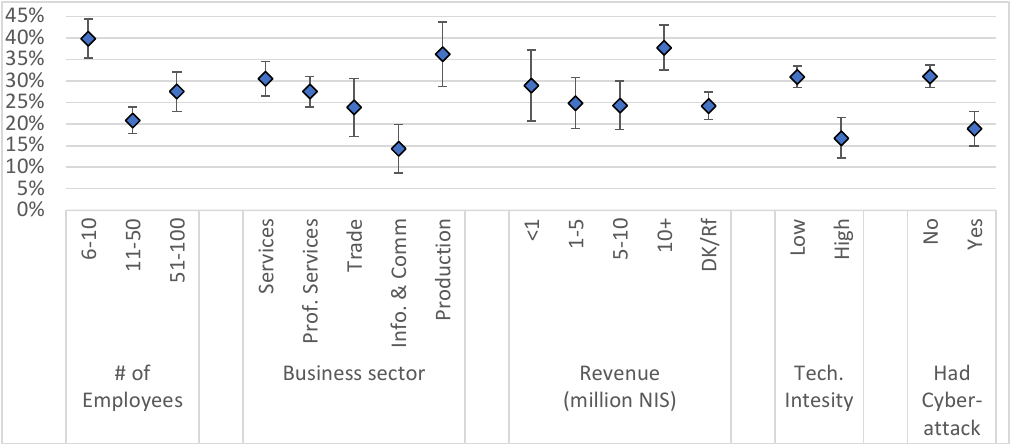}
        \caption{Low SA Level 3}
        \label{fig:aw3}
    \end{subfigure}
    \hfill
    \begin{subfigure}{0.47\textwidth}
        \includegraphics[width=\linewidth, trim=2 2 2 2,clip]{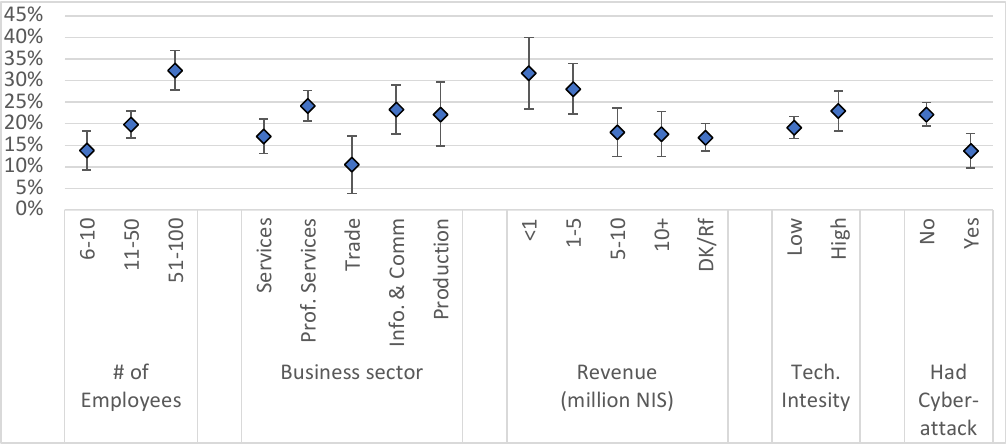}
        \caption{Low SA Level 4}
        \label{fig:aw4}
    \end{subfigure}
    \vspace{-3pt}
    \begin{subfigure}{0.47\textwidth}
        \includegraphics[width=\linewidth, trim=2 2 2 2,clip]{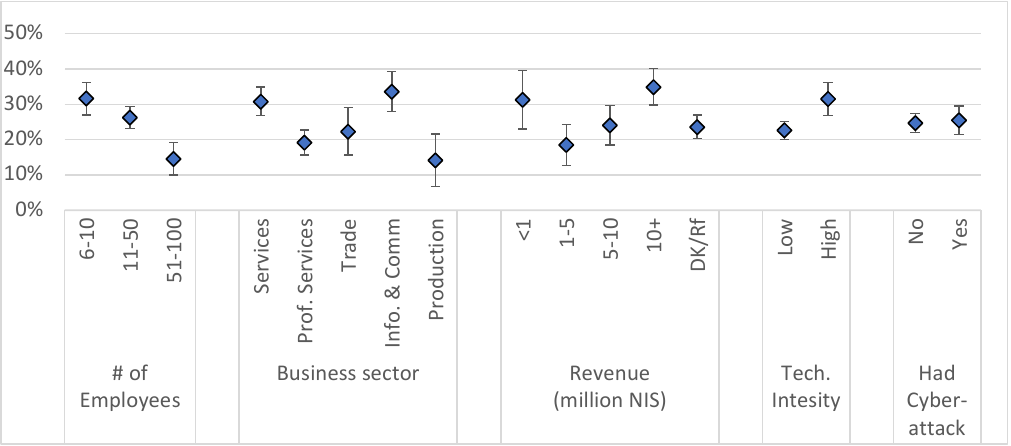}
        \caption{Low SA Level 5}
        \label{fig:aw5}
    \end{subfigure}
    \hfill
    \begin{subfigure}{0.47\textwidth}
        \includegraphics[width=\linewidth, trim=2 2 2 2,clip]{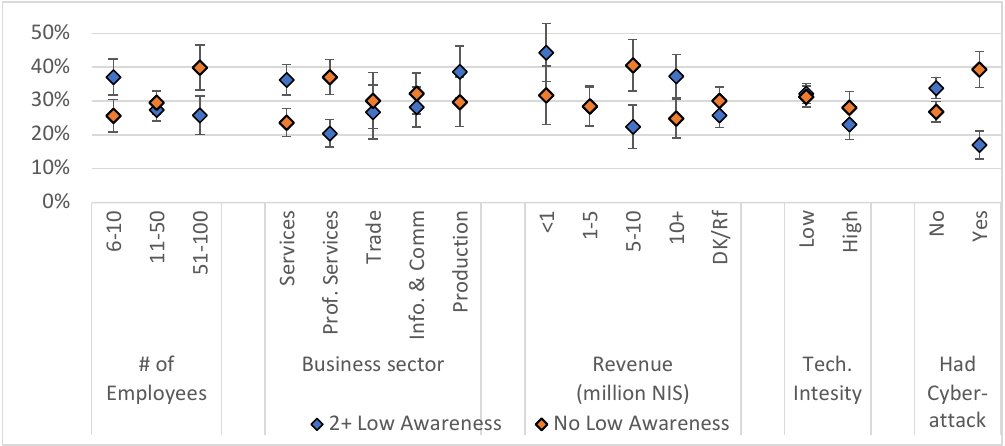}
        \caption{Multiple Low SA vs. No Low SA}
        \label{fig:aw2+}
    \end{subfigure}
    
    \caption{Margins from logistic regressions predicting the probabilities of decision-makers having low awareness.}
    \label{fig:aw}
    \vspace{-15pt}
\end{figure*}

\bsub{Digital Assets and Defense Distribution.}
Figure~\ref{fig:basic} shows participating SMB's current status in terms of valuable digital assets and deployed protective measures.
Based on our survey, the type of data that most SMBs own regardless of business attributes are personal data of the customers and employees. In addition, a majority of SMBs have websites available, and most use them as a means to communicate business information, such as for product viewing and service advertisements. For protective measures, 89\% of the SMBs claimed they define access permissions for individual employees. Specifically, every employee is assigned a username, and their security clearances are adjusted accordingly. This is followed by purchasing security solutions from third parties and practicing regular backup to cloud storage. Interestingly, we found that SMBs in Israel tend to choose technical measures (such as backups and access control) over training and simulations, which is similar to SMBs in Germany~\cite{huaman2021large}. Note that around 4\% and 8\% of the SMBs shared that they do not have any digital assets or protective measures, respectively.

\subsection{Awareness vs. Business Characteristics}

Figure~\ref{fig:aw} shows the marginal probabilities for low SA based on a logistic regression with business attributes (number of employees, business sector, annual revenue level, technological intensity, and cyber-attack experience). The corresponding coefficients, standard errors, and statistical significance are included in Appendix~\ref{app:awarenessprob}. We describe our findings below.

\bsub{Level 1.} 
Our analysis did not reveal statistically significant variables that are directly correlated with level 1 SA. However, we observed several interesting tendencies. We found that SMBs with less than 1M annual revenues are most likely to ignore the importance of cybersecurity, while SMBs that are in the Professional Service sector or have more employees can be more aware. Interestingly, those who have high technological intensity are more likely to be at low level 1 SA than others who have relatively lower technological intensity.

\bsub{Level 2.}
The average assessment given by all interviewees is 3.4, meaning a majority of decision-makers do not believe that they are easily exposed to attacks. Meanwhile, decision-makers in the Information and Communication sector are more aware of the risks of attacks ($\beta= -1.075, p < 0.05$), making them more prepared in case of cyber incidents. Decision-makers whose businesses have experienced attacks before generally perceive a higher risk of attacks than those who did not ($\beta= -1.19, p < 0.01$), indicating that decision-makers may learn from past experiences to raise awareness. In addition, businesses with 6-10 employees are more likely to overlook the risk of cyberattacks ($\beta= 0.576, p < 0.1$).

\bsub{Level 3.}
Based on our survey, more than half (54\%) of the respondents reported that they are familiar with official cybersecurity guidelines. Surprisingly, participants generally expressed a lower confidence score despite their claim on cyber guideline familiarity. This is especially evident in businesses with 6-10 employees, in which over half of the interviewees (53\%) claimed guideline familiarity but had an average confidence rating of only 3.8. This is also reflected in Figure~\ref{fig:aw3}, where businesses with less than 10 employees are the most likely to be at low awareness ($\beta= 1.023, p < 0.01$). Greater confidence in cybersecurity knowledge sufficiency is seen in those from the Information and Communication sector ($\beta= -1.045, p < 0.05$), and those from technology-intensive businesses ($\beta= -0.868, p < 0.05$). Experience with cyberattacks may also prompt decision-makers to understand security precautions more ($\beta= -0.720, p < 0.05$).

\bsub{Level 4.}
Referring to Figure~\ref{fig:aw4}, we see that SMBs from the Trade sector are the least likely to be ignorant about level 4 SA, though this is not statistically significant. We found that most decision-makers who are not aware of the high risk come from businesses with more than 50 employees ($\beta= 1.15, p < 0.05$). Businesses that have no prior attack experiences also tend to overlook the need to act ($\beta= -0.618, p < 0.1$).

\bsub{Level 5.}
Around one in five decision-makers reported that their available budgets and human resources prevent them from implementing better cyber precautions, while one in ten indicated the lack of time as a barrier. It is worth noting that businesses with more than 50 employees are more likely to experience resource shortage ($\beta= -0.775, p < 0.1$). As for the difference between economic sectors, the Professional Service sector ($\beta= -0.651, p < 0.1$) and the Production sector ($\beta= -1.026, p < 0.1$) are less likely to indicate a lack of resources. Revenue level, technological intensity, and past attack experiences do not significantly affect level 5 SA.

\bsub{Multiple Low Awareness.}
Figure~\ref{fig:aw2+} shows which type of business may have low SA on multiple levels.
Professional Service sector ($\beta= -0.857, p < 0.05$) and businesses that make 5M-10M NIS annually ($\beta= -1.103, p < 0.05$) are more likely to be aware of cybersecurity. 
Those who have experienced cyberattacks before are less likely to have multiple awareness issues ($\beta= -0.967, p < 0.01$).

\bsub{No Low Awareness.}
Figure~\ref{fig:aw2+} shows which type of business are likely to be free of any awareness issues. Businesses with more than 50 employees ($\beta= 0.729, p < 0.1$) and businesses with attack experiences ($\beta= 0.617, p < 0.05$) are likely to be more aware. Those with high technological intensity are less likely to have no low awareness issues ($\beta= -0.524, p < 0.1$).

\begin{tcolorbox}[compactbox]
    \textbf{\textit{Takeaway:}} Our results show that revenue level does not significantly impact individual situational awareness, though companies having 5-10 million NIS annually are less likely to have multiple awareness issues. We also see a positive impact of attacked experiences on awareness.
    In addition, being in the Information and Communication sector allows decision-makers to become familiar with awareness levels 2 and 3, while high technological intensity mainly increases level 3 awareness.
    Finally, a larger company size appears to negatively influence awareness level 4, possibly due to overconfidence or difficulty in management. Yet, it suggests sufficient resources to tackle security challenges. 
\end{tcolorbox}

\subsection{Holistic Structural Equation Model}
\begin{figure}[t]
    \centering
    \includegraphics[width=\columnwidth]{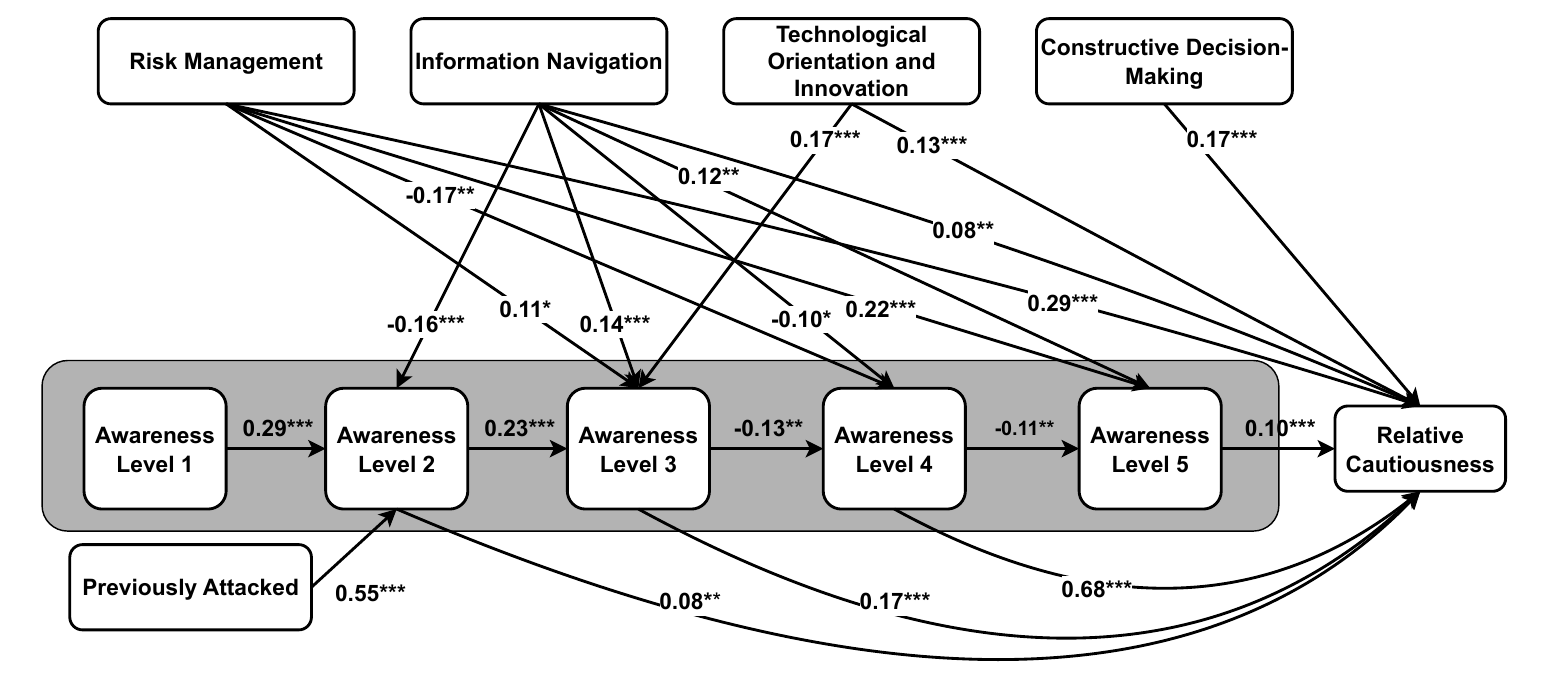}
    \vspace{-10pt}
    \caption{Structural equation model of situational awareness, root causes, and relative cautiousness.}
    \label{fig:sem}
\vspace{-15pt}
\end{figure}

We constructed a holistic structural equation model (SEM)~\cite{hoyle1995structural} from the collected data, showing the correlation of factors impacting key decision-maker's security awareness. The SEM draws relations between root causes, attack experience, relative cautiousness, and different levels of SA among all kinds of businesses, as shown in Figure~\ref{fig:sem}. Only the statistically significant arrows are shown.

\bsub{Correlation among SA Levels.} 
While level 1 is positively correlated with level 2, and level 2 is positively correlated with level 3, level 3 is negatively correlated with level 4. This implies that greater perceived knowledge of precautions can lead to false beliefs that there's no further need to act. In addition, level 4 is negatively correlated with level 5, indicating that resources such as time, budget, and personnel are essential and lacking for businesses that wish to actively defend against cyberattacks.
The estimated correlations between SA levels indicate that the SA model is a maturity model~\cite{mettler2011maturity}, as it is implied that each level influences the next, and thereby indirectly influences relative cautiousness. The model also suggests that levels 2, 3, and 4 directly influence relative cautiousness.

\bsub{Correlation with Other Factors.}
Except for awareness level 1, all SA levels are positively correlated with relative cautiousness. This shows that relative cautiousness can be improved by increasing SA. Moreover, experience with cyberattacks influences the perceived risk of attack exposure, which in turn motivates decision-makers to improve cybersecurity knowledge and readiness.
It is also shown that root causes are strongly related to SA and relative cautiousness. Therefore, they are critical when considering interventions for SMBs. We discuss the role of root causes in our model and the respective interventions in \S~\ref{subsec:rootResult}.
Besides the root causes' influence, SA is directly correlated with relative cautiousness. The correlation implies that other factors besides SA link root causes to security readiness. Such factors could be due to cultural tendencies or influences from other positions besides decision-makers, and should be further explored in future work.

\subsection{Impact of Root Causes}
\label{subsec:rootResult}

\bsub{Inadequate Risk Management.}
According to Table~\ref{tab:root}, most SMBs facing this obstacle are from the Trade and Service sector. Our holistic model suggested that risk management is highly correlated with SA and also directly correlated with relative cautiousness. Inadequate risk management affects decision-maker's level 3 and level 4 SA, making them under/over-estimate the urgency and failing to correctly allocate available resources. For these businesses, allocating resources effectively according to the situation is the key. 

\bsub{Difficulty in Information Navigation.}
Except for SA level 1, the ability to navigate abundant information can also impact SA and relative cautiousness. Note that the perception of information navigation difficulty is negatively correlated with SA level 2 and level 4, meaning the harder it is to navigate information, the more likely there is for an overestimation of risks. The average rating from decision-makers is 2.6, which is the lowest of the root causes. For these businesses, improved information sources are needed to mitigate biased awareness.

\bsub{Lack of Technological Orientation and Innovation.}
According to the holistic model, the lack of technological innovation directly affects SA level 3. This is also where decision-makers believe they need the most guidance. For these businesses, a way to facilitate discussion about available software or tactics is needed. The Information and Communication sector, whose businesses presumably engage more with technology, rates the highest in their technological knowledge (3.1). 

\bsub{Lack of Constructive Decision-Making.}
Finally, while constructive decision-making is related to a business's relative cautiousness, it does not seem to correlate with any of the SAs. According to Table~\ref{tab:root}, this is where decision-makers think they perform best and have the least issues comparatively.

\begin{table}[t]
\centering
\caption{Root Causes vs. Business Characteristics}
\label{tab:root}
\renewcommand{\arraystretch}{0.5}
\resizebox{\columnwidth}{!}{%
\scriptsize
\begin{tabular}{llcccc}
\specialrule{.2em}{.1em}{.1em}
& \textbf{} & \textbf{R.M.} & \textbf{I.N.} & \textbf{T.I.} & \textbf{C.D.M} \\ \specialrule{.2em}{.1em}{.1em}
\textbf{Total} &  & 3.0 & 2.6 & 2.6 & 3.6 \\ \hline
\multirow{2}{*}{\textbf{\begin{tabular}[c]{@{}l@{}}Business Size \\ (\# of Employees)\end{tabular}}} & 6-10 & 3.1 & 2.4 & 2.6 & 3.7 \\
 & 11-50 & 2.9 & 2.6 & 2.5 & 3.5 \\
\textbf{} & 51-100 & 3.3 & 2.9 & 2.8 & 3.8 \\ \hline
\textbf{Business Sector} & Services & 2.8 & 2.8 & 2.4 & 3.4 \\
\textbf{} & Prof. Services & 3.1 & 2.4 & 2.6 & 3.8 \\
\textbf{} & Trade & 2.7 & 2.3 & 2.7 & 3.6 \\
\textbf{} & Info. \& Comm. & 3.4 & 2.8 & 3.1 & 3.7 \\
\textbf{} & Production & 3.1 & 2.9 & 2.3 & 3.6 \\ \hline
\multirow{2}{*}{\textbf{\begin{tabular}[c]{@{}l@{}}Revenue \\ (million NIS)\end{tabular}}} & \textless{}1 & 2.9 & 2.6 & 2.6 & 3.7 \\
\textbf{} & 1-5 & 3.1 & 2.7 & 2.5 & 3.5 \\
\textbf{} & 5-10 & 2.8 & 2.8 & 2.7 & 3.2 \\
\textbf{} & 10+ & 3.8 & 2.4 & 2.6 & 3.6 \\
\textbf{} & Refuse to answer & 3.1 & 2.6 & 2.6 & 3.8 \\ \hline
\textbf{Technological Intensity} & Low & 3.0 & 2.6 & 2.5 & 3.6 \\
\textbf{} & High & 3.1 & 2.6 & 2.8 & 3.6 \\ \hline
\textbf{Experienced Cyberattack} & No & 3.0 & 2.6 & 2.6 & 3.6 \\
\textbf{} & Yes & 3.2 & 2.6 & 2.7 & 3.5 \\
\specialrule{.2em}{.1em}{.1em}
\end{tabular}%
}
    \begin{tablenotes}
    \scriptsize
    \item R.M.: Risk Management, I.N.: Information Navigation, T.I.: Technology Innovation, C.D.M: Constructive Decision Making
    \end{tablenotes}
\vspace{-15pt}
\end{table}

\section{Discussion}

\subsection{Answers to Research Questions}

With both qualitative and quantitative analysis, we have a glimpse of the security mindsets of SMB key decision-makers. Modeling SA into different levels and accounting for the root causes of low awareness shed light on the way SA could be improved. We finally answer our research questions.

\bsub{RQ1.}
We identified key decision-makers' perceived cyber threats based on the digital assets they valued. Alongside company data such as customer data, employee data, operational data, and intellectual property, many companies stated that they host websites for advertisement or customer communication, and are concerned with service disruption due to server downtime. Yet, comparing businesses with others showed that 23\% believed that they would not get attacked. To motivate security actions among SMBs, key decision-makers first need to be able to accurately evaluate their business's organizational risks. As prior attack experiences greatly influence risk perception, red team exercises may help guide the evaluation, identifying where false risk perceptions exist.

\bsub{RQ2.}
We observed a tendency for executives to mention more technical measures than employee training, which coincides with the findings from \cite{huaman2021large}. Interestingly, almost all participants unanimously agreed that backup to cloud storage is important for company operation. However, other forms of defense are mostly seen as redundant, with organizational measures receiving mixed preferences, and only minimal efforts are spent to enforce secure behaviors. 
We found 27\% of the businesses in our sample to have insufficient knowledge regarding security defenses. Like risk perceptions, accurate deployment and evaluation of defenses are also needed to enforce security cost-effectively, as the deployed defense may not necessarily mitigate risk and has its own set of harms. For instance, common industrial practice to combat phishing attacks was found to be ineffective by prior study~\cite{lain2022phishing}.
Figure~\ref{fig:sem} shows several ways this perception can be improved, including adequate risk management and getting familiar with both security knowledge and available defense technologies.

\bsub{RQ3.}
The capability to acknowledge and comprehend elements in the current situation to draw informative decisions marks level 2 maturity in Endsley's SA model.
Third-party consultants, lectures, news, and past security experiences are common information sources that SMB executives consider when making decisions. The impact of security incidents is two-sided. On the one hand, decision-makers would be motivated to implement more defense if the impact is costly. On the other, some would only view them as inconveniences and focus more on recovery but not defense. The status of other businesses may also influence how decision-makers perceive their own risks.
Meanwhile, whether the company has cyberattack experiences is a critical indicator of low SA regarding security. Businesses that have experienced attacks before are more likely to be free of any awareness issues. Having a larger business size seems to negatively affect decision-maker's awareness of the need to act, though resources are also more readily available. For economic sectors, businesses in the technology domain are usually more aware of the risks. For those who neglect the need to take cybersecurity actions, the issue can be addressed by assisting in risk management or easing information navigation, such as using channels familiar to decision-makers to convey concrete security suggestions.

\bsub{RQ4.}
We draw correlations between root causes and the level of awareness that they have an impact on. In general, decision-makers consider the navigation of security information and the adoption of innovative technology as major roadblocks in business operations. They perform averagely in terms of risk management, and consider the lack of constructive decision-making as a less critical problem. Except for SA level 1, these root causes can affect decision-maker's SA significantly.

\subsection{Interventions}
We aim to make our findings on SA and practical challenges broadly applicable despite this study being conducted in Israel. Though our results indicate universal problems faced by SMBs, the corresponding solutions for these issues may depend on a country's cultural, economic, and technological context. Based on this work, we have collaborated with Israeli government officials to devise actionable interventions to address the challenges and barriers accordingly.

\bsub{Networking and Institutional Guidance.}
Given that lectures were mentioned as one of the vital information sources, networking opportunities such as government-led conferences and workshops should be catered to provide educational lectures on how best to manage company resources, as well as providing a space to help build personal connections between SMB executives and government officials~\cite{billington2009effectiveness}. Since most SMBs facing the obstacle of inadequate risk management are from the Trade and Service sector (Table~\ref{tab:root}), policies regarding tax credits and guidance from financial institutions, such as the Small Business Lending Fund~\cite{DoT_2024}, may also help SMBs budget their capital.

\bsub{SMB-friendly Information Source.}
From the policy point of view, actionable standards and guidelines may help inform SMBs' legal obligation in matters of cybersecurity. In addition, a mix of information pulling and pushing leveraging intelligent agents, such as a central hub dedicated to the curation and sharing of cybersecurity knowledge, may be extremely useful in improving key decision-makers' experience during information navigation, helping them combat information overload~\cite{edmunds2000problem}. Subsidies on counseling services may offer extra aid in offloading some of the decision-making to dedicated information specialists~\cite{butcher1998meeting}.

\bsub{Identify Security Solutions through Technical Exchange.}
One way to tackle the issue with technological orientation is to host venues where merchants of security solutions can showcase their products. Such means of open-system orchestration help facilitate technical exchange between software companies, which can foster innovations that combat cyber criminals more effectively~\cite{dutt2016open,giudici2018open}. This will also allow SMB executives to understand what is currently available on the market, while letting them experience the products first-hand and communicate with the representatives about potential customization to fit their business.

\bsub{Preventive Assessment and Detection.}
It is recommended that decision-makers assess their business resiliency and find out potential vulnerabilities in advance, since organizations that practice regular security assessments experienced 40\% less unplanned downtime~\cite{CloudIBN}. There is also a need to encourage organizational measures such as employee training, emergency drills, or attack simulations. These help familiarize decision-makers with incidental situations and prepare them to make more informed decisions under urgency.
\section{Conclusion}
We conducted an initial semi-structured interview with 21 key decision-makers to understand what they consider when dictating a company's course of action regarding cybersecurity. Using the situational awareness model, we surveyed 322 key decision-makers to identify important factors influencing company executives' decision-making process, as well as find out the current awareness status of cybersecurity among Israeli SMBs. Based on our findings, we developed a holistic structural equation model considering potential root causes and relative cautiousness. In light of our results, we suggested interventions to overcome the identified barriers.

\bibliographystyle{plain} 
\bibliography{reference}{}

\appendix
\section{Interview Guide}
\label{app:interviewGuide}

\begin{enumerate}
    \small
    \setlength\itemsep{0em}
    \item Please start by telling me about yourself, including your education and familiarity with computer technology.
    \item Please tell me about your company, what it does, how long it has been operating, and the annual turnover.
    \item What kind of systems do you use and what information is stored? What is something you think has a high risk of losing and needs to be protected?
    \item Who is in charge of IT information security? If a third party is in charge, is there any specific reason that you hired him/them?
    \item What are the risks and consequences of your business being attacked? Have you heard talk of cyberattacks in your field?
    \item What are the protective measures that the company is using? Was there some cyber defense that you were unable to implement?
    \item Has the company experienced attacks before? What did you do after the attack?
    \item Can you share with me your sources of information for learning about cyber protection?
    \item Is there anything else you would like to share?
\end{enumerate}
\section{Survey Instrument}
\label{app:surveyinstrument}

\newcommand{\Qq}[1]{#1}

\newcommand{\QO}{$\bigcirc$}

\newcommand{\Qbox}{\hspace{9mm}\framebox[0.7\columnwidth]{\rule{0pt}{0.05cm}}}

\newcounter{qr}
\newcommand{\Qrating}[1]{\QO\forloop{qr}{1}{\value{qr} < #1}{---\QO}}

\newcommand{\Qline}[1]{\noindent\rule{#1}{0.6pt}}

\newcounter{ql}
\newcommand{\Qlines}[1]{\forloop{ql}{0}{\value{ql}<#1}{\vskip0em\Qline{\linewidth}}}

\newenvironment{Qlist}{%
\renewcommand{\labelitemi}{\QO}
\begin{itemize}[leftmargin=1.5em,topsep=-0.5em,itemsep=0em]
}{
\end{itemize}
}

\newlength{\qt}
\newcommand{\Qtab}[2]{
\setlength{\qt}{\linewidth}
\addtolength{\qt}{-#1}
\hfill\parbox[t]{\qt}{\raggedright #2}
}

\newcounter{itemnummer}
\newcommand{\Qitem}[2][]{
\ifthenelse{\equal{#1}{}}{\stepcounter{itemnummer}}{}
\ifthenelse{\equal{#1}{a}}{\stepcounter{itemnummer}}{}
\begin{enumerate}[topsep=2pt,leftmargin=2.8em]
\item[Q\arabic{itemnummer}#1.] #2
\end{enumerate}
}

\definecolor{bgodd}{rgb}{0.8,0.8,0.8}
\definecolor{bgeven}{rgb}{0.9,0.9,0.9}
\newcounter{itemoddeven}
\newlength{\gb}
\newcommand{\QItem}[2][]{
\setlength{\gb}{\linewidth}
\addtolength{\gb}{-5.25pt}
\ifthenelse{\equal{\value{itemoddeven}}{0}}{%
\noindent\colorbox{bgeven}{\hskip-3pt\begin{minipage}{\gb}\Qitem[#1]{#2}\end{minipage}}%
\stepcounter{itemoddeven}%
}{%
\noindent\colorbox{bgodd}{\hskip-3pt\begin{minipage}{\gb}\Qitem[#1]{#2}\end{minipage}}%
\setcounter{itemoddeven}{0}%
}
}
\scriptsize

\bsub{Screening}
\Qitem{ \Qq{Which of the following best describes your business ownership?}
\begin{Qlist}
\vspace{-\baselineskip}
    \begin{multicols}{3}
        \item Privately Owned
        \item Publicly Owned 
        \item Cooperative Owned
        \item Non-profit 
        \item Government Owned
    \end{multicols}
\vspace{-1.5\baselineskip}
\end{Qlist}
}

\Qitem{ \Qq{How many employees are in your business?}
}

\Qitem{ \Qq{What is your position in the business? (Select all that apply)}
\begin{Qlist}
\vspace{-\baselineskip}
    \begin{multicols}{3}
        \item Business Owner 
        \item Vice President
        \item CEO 
    \end{multicols}
\vspace{-3\baselineskip}
    \begin{multicols}{2}
        \item Department/Division Manager
        \item Non-management Role
    \end{multicols}
\vspace{-1.5\baselineskip}    
\end{Qlist}
}

\Qitem{ \Qq{What type of business do you own?}
}

\Qitem{ \Qq{What is the economic sector of the business?}
\begin{Qlist}
    \item Activities in real estate (Service)
    \item Management and support services (Service)
    \item Wholesale and retail trade and repair of motor vehicles (Trade)
    \item Industry, mining and quarrying (Production)
    \item Electricity and water supply, sewage services and waste treatment (Service)
    \item Professional, scientific and technical services (Professional Service)
    \item Information and communication 
    \item Hospitality and food services (Service)
    \item Transportation, storage, mail and courier services (Service)
    \item Financial services and insurance services (Professional Service)
\vspace{-1.3\baselineskip}
    \begin{multicols}{2}
        \item Other:\_\_\_\_ 
        \item DK/RF 
    \end{multicols}
\vspace{-1.5\baselineskip}
\end{Qlist}
}

\Qitem{ \Qq{What was your company’s annual revenue (in NIS) for 2022? Your answers will not be transferred to any business entity.}
\begin{Qlist}
\vspace{-\baselineskip}
\begin{multicols}{3}
    \item Up to 1 million
    \item 1-5 million 
    \item 5-10 million 
\end{multicols}
\vspace{-2.7\baselineskip}
\begin{multicols}{3}
    \item 10-50 million 
    \item 50+ million 
    \item DK/RF
\end{multicols}
\vspace{-1.5\baselineskip}
\end{Qlist}
}


\bsub{Business Background}
\Qitem{ \Qq{In what year was your business established?}
}

\Qitem{ \Qq{How many of your employees use a computer when they work?}
}

\Qitem{ \Qq{Are there standards and/or regulations for information security that your company implements?}
\begin{Qlist}
\vspace{-\baselineskip}
    \begin{multicols}{3}
        \item Yes, please specify
        \item No
        \item DK/RF
    \end{multicols}
\vspace{-1.5\baselineskip}
\end{Qlist}
}

\Qitem{ \Qq{Does the business operate outside of Israel?}
\begin{Qlist}
\vspace{-\baselineskip}
    \begin{multicols}{2}
        \item Yes
        \item No
    \end{multicols}
\vspace{-1.5\baselineskip}
\end{Qlist}
}

\Qitem{ \Qq{Where is your business located?}
\begin{Qlist}
\vspace{-\baselineskip}
    \begin{multicols}{2}
        \item Located at one site
        \item Located at several sites
    \end{multicols}
\vspace{-1.5\baselineskip}    
\end{Qlist}}

\Qitem{ \Qq{Are you a member of a business association?}
\begin{Qlist}
\vspace{-\baselineskip}
    \begin{multicols}{3}
        \item Yes, please specify
        \item No
        \item DK/RF
    \end{multicols}
\vspace{-1.5\baselineskip}
\end{Qlist}
}

\Qitem{ \Qq{In which city do most of your business’s activity take place?}
}


\bsub{Risk Exposure}
\Qitem{ \Qq{Does your business have an employee who is in charge of computing?}
\begin{Qlist}
\vspace{-\baselineskip}
    \begin{multicols}{3}
        \item Yes
        \item No
        \item DK/RF
    \end{multicols}
\vspace{-1.5\baselineskip}
    \item No, an external party/person provides my business with computing services
\end{Qlist}
}

\Qitem{ \Qq{\textcolor{red}{[If Q14 == Yes]} Who is in charge of overseeing all aspects of computing matters for your business, including information security?}
\begin{Qlist}
\vspace{-\baselineskip}
    \begin{multicols}{3}
        \item CIO 
        \item ICT 
        \item CISO
    \end{multicols}
\vspace{-2.7\baselineskip}
    \begin{multicols}{3}
        \item Other:\_\_\_\_
        \item DK/RF
    \end{multicols}
\vspace{-1.5\baselineskip}
\end{Qlist}
}

\Qitem{ \Qq{\textcolor{red}{[If Q14 == external party/person]} How would you best describe the relationship you have with your computing services company/person? (Select all that apply)}
\begin{Qlist}
    \item We are in touch when there are technical problems.
    \item We hold regular periodic meetings at least once a year.
    \item We receive from him/them general updates on new technologies.
    \item We receive recommendations from them to buy cybersecurity products.
\vspace{-1.3\baselineskip}
    \begin{multicols}{3}
        \item Other:\_\_\_\_
        \item DK/RF 
    \end{multicols}
\vspace{-1.5\baselineskip}
\end{Qlist}
}

\Qitem{ \Qq{Do the employees in your business have the choice of working remotely?}
\begin{Qlist}
    \item Yes, all of our employees can work remotely.
    \item Yes, some of our employees can work remotely.
\vspace{-1.3\baselineskip}
    \begin{multicols}{3}
        \item No
        \item DK/RF
    \end{multicols}
\vspace{-1.5\baselineskip}
\end{Qlist}
}

\Qitem{ \Qq{Is software installed on local servers in the business or are they on the cloud?}
\begin{Qlist}
    \item All our software is on the cloud only
    \item Some of the software is located on the cloud and some on local servers
    \item All our software is located local server only
    \item DK/RF
\end{Qlist}
}

\Qitem{ \Qq{Does your business use Customer relationship management (CRM)?}
\begin{Qlist}
\vspace{-\baselineskip}
    \begin{multicols}{4}
        \item Local CRM 
        \item Cloud CRM
        \item No 
        \item DK/RF
    \end{multicols}
\vspace{-1.5\baselineskip}
\end{Qlist}
}

\Qitem{ \Qq{Does your business use Enterprise resource planning (ERP)?}
\begin{Qlist}
\vspace{-\baselineskip}
    \begin{multicols}{4}
        \item Local ERP 
        \item Cloud ERP
        \item No 
        \item DK/RF
    \end{multicols}   
\vspace{-1.5\baselineskip}
\end{Qlist}
}

\Qitem{ \Qq{Does your business have a website?}
\begin{Qlist}
\vspace{-\baselineskip}
    \begin{multicols}{4}
        \item Yes
        \item No
        \item DK/RF
    \end{multicols}
\vspace{-1.5\baselineskip}
\end{Qlist}
}

\Qitem{ \Qq{\textcolor{red}{[If Q21 == Yes]} How is the website managed?}
\begin{Qlist}
    \item Independent management from our business.
    \item Webpage on other websites such as Amazon, Etsy, Ebay, etc.
\vspace{-1.3\baselineskip}
    \begin{multicols}{2}
        \item Other:\_\_\_\_ 
        \item DK/RF
    \end{multicols}
\vspace{-1.5\baselineskip}
\end{Qlist}
}

\Qitem{ \Qq{\textcolor{red}{[If Q21 == Yes]} What is the purpose of the website? (Select all that apply)}
\begin{Qlist}
    \item For business information: viewing products/services offered by the business
    \item For selling products or services, and charging the customer for the purchase
    \item For individualized use, where each user can sign in with a personal account 
    \item The service provided by the business is located on the website (SaaS)
\vspace{-1.3\baselineskip}
    \begin{multicols}{3}
        \item Other
        \item DK/RF
    \end{multicols}
\vspace{-1.5\baselineskip}
\end{Qlist}
}

\Qitem{ \Qq{How are the payments completed?}
\begin{Qlist}
\vspace{-1.3\baselineskip}
    \begin{multicols}{2}
        \item Charged directly on our website
        \item Charged on an external website
        \item Payment applications
        \item Bank transfer 
    \end{multicols}
\vspace{-2.7\baselineskip}    
    \begin{multicols}{3}
        \item Check/Cash
        \item Other
        \item DK/RF
    \end{multicols}
\vspace{-1.5\baselineskip}
\end{Qlist}
}

\Qitem{ \Qq{Which of the following digital assets does your business have? (Select all that apply)}
\begin{Qlist}
    \item Customer data (customer names, personal details)
    \item Customer financial information (credit cards, bank accounts, etc.) 
    \item Customer sensitive data (medical information) 
    \item Employee data (personal data, shifts, salaries, etc.)
    \item Operational data (details pertaining to machines, materials, etc.)
    \item Intellectual property (software projects, engineering plans, etc.)
    
\vspace{-1.3\baselineskip}
    \begin{multicols}{2}
    \item Company financial data
    \item DK/RF
    \end{multicols}
\vspace{-1.5\baselineskip}    
\end{Qlist}
}

\Qitem{ \Qq{A cyberattack has the potential to harm the digital assets of the business, including their destruction or theft. Please assess the severity of potential damage or loss for each of the digital assets on a scale of 1 (minimal damage) to 10 (most significant damage). \textit{[Use list of digital assets selected in the previous question.]}}
}


\bsub{Situational Awareness}
\Qitem{ \Qq{In your opinion, what is the greatest possible damage that could occur in the event of the loss or theft of all the digital assets of your business?}
\begin{Qlist}
    \item Bankruptcy 
    \item A significant decrease in income/revenue
    \item There will be a cost to restore the information 
    \item There will be a decrease in business productivity 
    \item Harm to the motivation levels of the business team 
\vspace{-1.3\baselineskip}
    \begin{multicols}{2}
    \item Harm to the business reputation
    \item Fines 
    \item Other:\_\_\_\_
    \item There will be no damage/harm
    \end{multicols}
\vspace{-1.5\baselineskip}        
\end{Qlist}
}

\Qitem{ \Qq{In your estimation, what is the likelihood that a business like yours will be attacked in the next year? On a scale of 1 to 10, with 1 indicating not likely at all, and 10 indicating extremely likely that an attack will occur.}
}

\Qitem{ \Qq{Do you know the guidelines on cyber-related issues from official sources in Israel and worldwide?}
\begin{Qlist}
\vspace{-1.5\baselineskip}
    \begin{multicols}{2}
        \item Yes, fully
        \item Yes, partially
        \item No
        \item Refuse to answer
    \end{multicols}
\vspace{-1.5\baselineskip}
\end{Qlist}
}

\Qitem{ \Qq{What are your sources of information in cybersecurity? (Select all that apply)}
\begin{Qlist}
\vspace{-\baselineskip}
    \begin{multicols}{2}
        \item Newsletter and magazines
        \item Lectures and conferences
    \end{multicols}
\vspace{-1.5\baselineskip} 
    \item Conversations with colleagues, other business owners 
    \item Government websites (Agency for Small and Medium Businesses, the National Cyber Array, etc.)
\vspace{-1.3\baselineskip}     
    \begin{multicols}{2}
        \item Internet forums
        \item Other:\_\_\_\_
    \end{multicols}
\vspace{-1.5\baselineskip}    
\end{Qlist}
}

\Qitem{ \Qq{To what extent do you think the knowledge you have in the field of cybersecurity is sufficient? On a scale of 1 to 10, with 1 indicating not at all sufficient, and 10 indicating extremely sufficient.}
}

\Qitem{ \Qq{How does your business protect itself from cyberattacks? (Select all that apply)}
\begin{Qlist}
    \item Purchasing security products (antivirus, firewall, and more) 
    \item Everyone has a username and their security settings are adjusted accordingly
    \item Implementing information security procedures
    \item Compliance with information security standards and authorization as a regulatory requirement
    \item Conducting penetration tests by external parties
\vspace{-1.3\baselineskip}
    \begin{multicols}{2}
        \item Requiring a password
        \item Information security training
        \item Incident response plan
        \item Routine risk assessment
        \item Emergency drills
        \item Phishing simulation    
        \item Employee training
        \item Keeping all software up to date
        \item Local backup 
        \item Cloud backup
        \item Other:\_\_\_\_ 
        \item Don't know  
    \end{multicols}
\vspace{-1.5\baselineskip}
\end{Qlist}
}

\Qitem{ \Qq{To the best of your knowledge, what is the level of cyber protection in your business? On a scale of 1 to 10, with 1 indicating a very low level of cybersecurity protection and 10 indicating a very high level of cybersecurity protection.}
}

\Qitem{ \Qq{What are the reasons you chose this score?}
}

\Qitem{ \Qq{What is the maximum amount in NIS you would be willing to invest annually to ensure cybersecurity measures?}
\begin{Qlist}
\vspace{-\baselineskip}
    \begin{multicols}{2}
        \item No need to invest at all
        \item up to 5,000
        \item 5,000-10,000
        \item 10,000-20,000
        \item 20,000-50,000
        \item 50,000-100,000
        \item Over 100,000
    \end{multicols}
\vspace{-1.5\baselineskip}
\end{Qlist}
}

\Qitem{ \Qq{In your opinion, what is the annual budget in NIS that a business like yours should invest in cybersecurity?}
\begin{Qlist}
\vspace{-\baselineskip}
    \begin{multicols}{2}
        \item No need to invest at all
        \item up to 5,000
        \item 5,000-10,000
        \item 10,000-20,000
        \item 20,000-50,000
        \item 50,000-100,000
        \item Over 100,000
    \end{multicols}
\vspace{-1.5\baselineskip}
\end{Qlist}
}

\Qitem{ \Qq{In your opinion, does the business invest enough budget for cybersecurity?}
\begin{Qlist}
\vspace{-\baselineskip}
    \begin{multicols}{2}
        \item Much more than necessary
        \item A little more than necessary
    \end{multicols}
\vspace{-1.5\baselineskip}        
        \item Approximately the amount needed
\vspace{-1.3\baselineskip}
    \begin{multicols}{2}
        \item A little less than necessary
        \item Much less than necessary
    \end{multicols}
\vspace{-1.5\baselineskip}
        \item DK/RF
\end{Qlist}
}

\Qitem{ \Qq{In your opinion, how many monthly hours (meetings, reading material, consultations, etc.) should a manager like you devote to cybersecurity?}
\begin{Qlist}
\vspace{-\baselineskip}
    \begin{multicols}{2}
        \item No need to spend time at all
        \item up to 5 hours
        \item 5-10
        \item 10-20
        \item 20-30
        \item 30-50
        \item More than 50
        \item DK/RF
    \end{multicols}
\vspace{-1.5\baselineskip}
\end{Qlist}
}

\Qitem{ \Qq{Are you devoting enough time to cybersecurity?}
\begin{Qlist}
\vspace{-\baselineskip}
    \begin{multicols}{2}
        \item Much more than necessary
        \item A little more than necessary
    \end{multicols}
\vspace{-1.5\baselineskip}        
        \item Approximately the amount needed
\vspace{-1.3\baselineskip}
    \begin{multicols}{2}
        \item A little less than necessary
        \item Much less than necessary
    \end{multicols}
\vspace{-1.5\baselineskip}
        \item DK/RF
\end{Qlist}
}

\Qitem{ \Qq{Please indicate whether you agree with the following statements. On a scale of 1 to 5, with 1 indicating strongly disagree and 5 indicating strongly agree.}
\begin{Qlist}
    \item My competitors have implemented or are in the process of implementing cybersecurity measures.
    \item My customers want my business to implement cybersecurity measures.
    \item Businesses I interact with believe we need to adopt cybersecurity measures.
\end{Qlist}
}

\Qitem{ \Qq{Has your business experienced cyberattacks?}
\begin{Qlist}
\vspace{-\baselineskip}
    \begin{multicols}{2}
    \item No
    \item Yes, once in the last year
    \item Yes, several times in the last year
    \item Yes, more than a year ago
    \item DK/RF
    \end{multicols}
\vspace{-1.5\baselineskip}
\end{Qlist}
}

\Qitem{ \Qq{Has your business faced the following due to security problems?} 
\begin{Qlist}
    \item Attempt to cause unavailability of the information and communication systems (such as ransomware)
    \item Attempt to cause destruction or corruption of information
    \item Attempt to cause disclosure of confidential data (e.g. phishing) 
\end{Qlist}
}

\Qitem{ \Qq{What was the extent of the damage? (Select all that apply)} 
\begin{Qlist}
\vspace{-\baselineskip}
    \begin{multicols}{2}
        \item No damage at all
        \item Ransom payment
        \item Money for additional computing services
        \item Damage to hardware
        \item Damage to reputation
        \item Damage to employee morale
        \item Man-hours for fixing
        \item Other:\_\_\_\_
        \item DK/RF
    \end{multicols}
\vspace{-1.5\baselineskip}
\end{Qlist}
}

\Qitem{ \Qq{Do you know of a business that experienced a cyber-attack? (Select all that apply)}
\begin{Qlist}
    \item Yes, a close colleague/acquaintance of mine experienced a cyber-attack
    \item Yes, I heard business in the same sector as mine experienced cyber-attack
    \item Yes, there are businesses that I do not know personally being attacked.  
    \item I never heard of cyberattacks occurring to others. 
    \item Refuse to answer
\end{Qlist}
}

\Qitem{ \Qq{How much do the following statements limit your implementation of cyber defense measures in your business? On a scale of 1 to 5, with 1 indicating limits very much and 5 indicating does not limit at all.}
\begin{Qlist}
    \item I have no contact with a security expert
    \item No clear instructions from reliable sources regarding the required actions
    \item I don't have a suitable technological understanding
    \item The employees are not involved in this matter
    \item The management team is not involved in this matter
    \item Lack of budget to implement the guidelines
    \item There is a lack of personnel who can implement the guidelines
    \item I have no one to consult in my social circle
    \item I have no time
\end{Qlist}
}

\Qitem{ \Qq{Does your business hold executive meetings regarding cybersecurity?}
\begin{Qlist}
\vspace{-\baselineskip}
    \begin{multicols}{2}
        \item Never
        \item Once a year or less
    \end{multicols}
\vspace{-1.3\baselineskip}
        \item More than once a year - once every quarter
        \item At least once a quarter – once a month 
\vspace{-1.3\baselineskip}
    \begin{multicols}{2}
        \item More than once a month 
        \item Refuse to answer 
    \end{multicols}
\vspace{-1.5\baselineskip}
\end{Qlist}
}

\Qitem{ \Qq{In the case that you would want to implement new cybersecurity guidelines that will require changing work habits, to what extent do you think the employees will cooperate in implementing the guidelines?}
\begin{Qlist}
\vspace{-\baselineskip}
    \begin{multicols}{3}
        \item Extremely
        \item Very much
        \item Slightly 
        \item Not at all
        \item Refuse to answer
    \end{multicols}
\vspace{-1.5\baselineskip}
\end{Qlist}
}


\bsub{Root Causes}
\Qitem{ \Qq{Which of the following statements best conveys your tendency to act when it comes to implementing new technologies in the business?}
\begin{Qlist}
    \item New technology is implemented in the business only if the existing technology is no longer possible
    \item New technology is implemented in the business only if it has an external demand from customers, suppliers, or regulators.
    \item New technology is implemented in the business only after we see that it proves itself in businesses similar to mine
    \item We strive to be ahead of our competitors when it comes to implementing new technologies that have just been released
\end{Qlist}
}

\Qitem{ \Qq{The following statements refer to your personal attitudes regarding cybersecurity. There are no right or wrong answers. Please provide your opinion on the following statements using a scale of 1 to 5, with 1 indicating strongly disagree and 5 indicating strongly agree.}
\begin{Qlist}
    \item Cybersecurity is an important issue that should concern all businesses. 
    \item My business is at risk of experiencing a cyber-attack.
    \item Cybersecurity threats are constantly evolving, so it's hard to stay up-to-date.
    \item I believe that the existing cybersecurity measures implemented in the business effectively safeguard against cyberattacks.
    \item I believe that cybersecurity measures are too expensive and are not worth the investment.
\end{Qlist}
}

\Qitem{ \Qq{During the last three years, has your business invested any resources in exploring new ideas for innovation? (For example, through participation in conferences, fairs, or exhibitions, following scientific/technical journals or commercial publications, information from professional organizations, social networks, or online business platforms)}
\begin{Qlist}
\vspace{-\baselineskip}
    \begin{multicols}{2}
        \item Did not invest resources at all
        \item Invested few resources
    \end{multicols}
\vspace{-1.5\baselineskip}        
        \item Invested a moderate amount of resources
\vspace{-1.3\baselineskip}
    \begin{multicols}{2}
        \item Invested several resources
        \item Invested much resources
    \end{multicols}
\vspace{-1.5\baselineskip}
        \item DK/RF
\end{Qlist}
}

\Qitem{ \Qq{To what extent is your business exposed to information about innovations made by similar companies? (Information regarding product development, production technologies, marketing methods, etc.)}
\begin{Qlist}
\vspace{-\baselineskip}
    \begin{multicols}{2}
        \item Not exposed to this information
        \item Exposed a little
        \item Exposed moderately
        \item Exposed to a great extent
        \item Extremely exposed
        \item DK/RF
    \end{multicols}    
\vspace{-1.5\baselineskip}
\end{Qlist}
}

\Qitem{ \Qq{The following set of questions are related to the ways in which you make decisions. There are no right or wrong answers. Please rate how strongly you agree with the following statements on a scale from 1 (strongly disagree) to 5 (strongly agree).}
\begin{Qlist}
    \item We rely on the personal experience of the management team
    \item We rely on the experience of the employees in the organization
    \item We rely on intuition and gut feelings
    \item We rely on information from external consultants 
    \item We rely on data, facts, and insights
\end{Qlist}
}

\Qitem{ \Qq{Does your business implement a risk management program?}
\begin{Qlist}
\vspace{-\baselineskip}
\begin{multicols}{3}
    \item Yes
    \item No
    \item DK/RF
\end{multicols}
\vspace{-1.5\baselineskip}
\end{Qlist}
}

\Qitem{ \Qq{On a scale from 1 (strongly disagree) to 5 (strongly agree), please indicate your level of agreement with the following statements.}
\begin{Qlist}
    \item We have a clear understanding of the risks the business can face
    \item We take actions to reduce risks
    \item We have contingency plans in the case that potential risks actually do occur
    \item Other issues in business management take priority over risk management
\end{Qlist}
}

\Qitem{ \Qq{Which of the following types of insurance does your business have? (Select all that apply)}
\begin{Qlist}
\vspace{-\baselineskip}
\begin{multicols}{2}
    \item Building insurance
    \item Content insurance
    \item Third-party insurance
    \item Professional liability insurance
    \item Employers liability insurance
    \item Product liability insurance
    \item Loss of profits insurance
    \item  Cyber insurance
    \item Other:\_\_\_\_
    \item DK/RF
\end{multicols}
\vspace{-1.5\baselineskip}
\end{Qlist}
}

\Qitem{ \Qq{The following statements refer to your attitudes regarding cybersecurity. There are no right or wrong answers. Please rate the following from 1 (strongly disagree) to 5 (strongly agree).}
\begin{Qlist}
    \item cyberattacks are a growing threat to businesses.
    \item My business is too small for hackers to bother attacking it.
    \item There is too much information circulating around cyberattacks that it overwhelms and confuses me.
    \item My business was not attacked so what we are doing is probably good enough.
    \item Small and medium-sized companies do not have the means to follow and implement all the guidelines in the field of cybersecurity. 
\end{Qlist}
}

\bsub{Interviewee Demographics}
\Qitem{ \Qq{How old are you? (in years)}
}

\Qitem{ \Qq{What is your gender?}
\begin{Qlist}
\vspace{-\baselineskip}
\begin{multicols}{4}
    \item Male
    \item Female
    \item Other:\_\_\_\_
    \item RF    
\end{multicols}
\vspace{-1.5\baselineskip}
\end{Qlist}
}

\Qitem{ \Qq{What is the highest level of education you completed?}
\begin{Qlist}
\vspace{-\baselineskip}
\begin{multicols}{2}
    \item Primary or middle school graduation certificate 
    \item Matriculation (without certificate)
    \item Matriculation certificate 
    \item Vocational certificate (secondary studies) 
    \item Certificate that is not an academic degree such as technician or engineer 
    \item Bachelor's degree or equivalent
    \item Master's degree or equivalent, including M.D. 
    \item Ph.D. or equivalent
    \item Yeshiva
    \item Other:\_\_\_\_
    \item Refuse to answer    
\end{multicols}
\vspace{-1.5\baselineskip}
\end{Qlist}
}

\Qitem{ \Qq{How long have you held your \textit{current position} in the business?}
}

\Qitem{ \Qq{How long have you been in this profession? }
}

\Qitem{ \Qq{How would you describe your level of technological knowledge?}
\begin{Qlist}
    \item No knowledge: I don't use a computer.
    \item Basic knowledge: I can use a computer for basic purposes, such as working with Microsoft Word.
    \item Intermediate level of knowledge: I feel comfortable using a computer and can solve problems on my computer if necessary.
    \item Advanced: I have advanced ability to install programs/solve related problems.
    \item Professional: I have professional background and the ability to program; professional knowledge of advanced technologies; relevant formal education
    \item Refuse to answer    
\end{Qlist}
}

\Qitem{ \Qq{Where did you acquire your technological knowledge and skills?
(Select all that apply)}
\begin{Qlist}
\vspace{-\baselineskip}
\begin{multicols}{2}
    \item I never acquired technological skills
    \item High school studies/engineer
    \item Academia (Bachelor's and Master's degrees)
    \item Professional training
    \item Military service
    \item Work experience
    \item Personal experience / Self-taught
    \item Other:\_\_\_\_
    \item Refuse to answer  
\end{multicols}
\vspace{-1.5\baselineskip}
\end{Qlist}
}
\section{Expected Damage from Cyber-Attacks from Interview Participants}

\begin{table}[h]
\centering
\setlength\tabcolsep{0pt}
\renewcommand{\arraystretch}{0}
\scriptsize
\begin{tabular*}{0.9\columnwidth}{@{\extracolsep{\fill}}cccccccc}
\specialrule{.2em}{.1em}{.1em}
\multicolumn{1}{c}{\textbf{\#}} & \multicolumn{1}{l}{\textbf{\rot{\parbox{1.5cm}{Data \\Recovery Cost}}}} & \textbf{\rot{\parbox{1.5cm}{*Operational \\Damage}}} & \multicolumn{1}{l}{\textbf{\rot{\parbox{1.5cm}{*Financial \\Fines}}}} & \multicolumn{1}{l}{\textbf{\rot{\parbox{1.5cm}{IP Leakage}}}} & \multicolumn{1}{l}{\textbf{\rot{\parbox{1.5cm}{Reputational \\Damage}}}} & \multicolumn{1}{l}{\textbf{\rot{\parbox{1.5cm}{Bankruptcy}}}} & \textbf{\rot{\parbox{1.5cm}{Aggregated \\Damage \\Severity}}} \\ \specialrule{.2em}{.1em}{.1em}
\multicolumn{1}{c}{P1} & \checkmark & \checkmark &  &  &  &  & \like{2} \\
\multicolumn{1}{c}{P2} & \checkmark &  &  &  &  &  & \like{0} \\
\multicolumn{1}{c}{P3} & \checkmark &  &  &  &  &  & \like{0} \\
\multicolumn{1}{c}{P4} & \checkmark & \checkmark & \checkmark &  & \checkmark & \checkmark & \like{6} \\
\multicolumn{1}{c}{P5} & \checkmark & \checkmark & \checkmark &  & \checkmark & \checkmark & \like{6} \\
\multicolumn{1}{c}{P6} & \checkmark &  & \checkmark & \checkmark &  &  & \like{4} \\
\multicolumn{1}{c}{P7} &  & \checkmark & \checkmark & \checkmark & \checkmark & \checkmark & \like{6} \\
\multicolumn{1}{c}{P8} & \checkmark &  &  &  &  &  & \like{0} \\
\multicolumn{1}{c}{P9} &  &  &  & \checkmark &  &  & \like{4} \\
\multicolumn{1}{c}{P10} & \checkmark & \checkmark & \checkmark &  &  &  & \like{4} \\
\multicolumn{1}{c}{P11} &  &  & \checkmark & \checkmark & \checkmark & \checkmark & \like{6} \\
\multicolumn{1}{c}{P12} &  & \checkmark & \checkmark &  & \checkmark &  & \like{4} \\
\multicolumn{1}{c}{P13} & \checkmark & \checkmark & \checkmark &  & \checkmark &  & \like{4} \\
\multicolumn{1}{c}{P14} & \checkmark &  &  &  &  &  & \like{0} \\
\multicolumn{1}{c}{P15} & \checkmark & \checkmark & \checkmark &  & \checkmark &  & \like{4} \\
\multicolumn{1}{c}{P16} & \checkmark & \checkmark & \checkmark & \checkmark & \checkmark & \checkmark & \like{8} \\
\multicolumn{1}{c}{P17} & \checkmark & \checkmark & \checkmark &  & \checkmark & \checkmark & \like{6} \\
\multicolumn{1}{l}{P18} & \checkmark & \checkmark &  &  &  & \checkmark & \like{4} \\
\multicolumn{1}{l}{P19} & \checkmark & \checkmark & \checkmark & \checkmark & \checkmark & \checkmark & \like{8} \\
\multicolumn{1}{l}{P20} & \checkmark & \checkmark & \checkmark & \checkmark & \checkmark & \checkmark & \like{8} \\
\multicolumn{1}{l}{P21} & \checkmark & \checkmark &  & \checkmark &  &  & \like{4} \\ \specialrule{.2em}{.1em}{.1em}
\end{tabular*}%
    \begin{tablenotes}
        \scriptsize
        \item *Operational Damage: No access to the business computers temporarily
        \item *Financial Fines: Due to failure to comply with regulations 
        \item \like{8} - Very Highly, \like{6} - Highly, \like{4} - Medium, \like{2} - Low, \like{0} - Very Low
    \end{tablenotes}
\vspace{-10pt}
\end{table}
\onecolumn

\section{Interview Qualitative Analysis Codebook}
\label{app:codebook}

\begin{table}[h]
\centering
\resizebox{\textwidth}{!}{%
\begin{tabular}{|c|cc|l|c|c|l|l|}
\hline
\multirow{9}{*}{\begin{tabular}[c]{@{}c@{}}\textbf{Digital} \\ \textbf{Asset}\end{tabular}} & \multicolumn{2}{c|}{\multirow{3}{*}{Customer data}} & list of customers & \multirow{9}{*}{\begin{tabular}[c]{@{}c@{}}\textbf{Awareness} \\ \textbf{Impact}\end{tabular}} & \multirow{3}{*}{Risk covered by} & consulting agency & \multirow{9}{*}{} \\ \cline{4-4} \cline{7-7}
 & \multicolumn{2}{c|}{} & financial data &  &  & insurance &  \\ \cline{4-4} \cline{7-7}
 & \multicolumn{2}{c|}{} & other sensitive data &  &  & government agency &  \\ \cline{2-4} \cline{6-7}
 & \multicolumn{2}{c|}{\multirow{2}{*}{Business data}} & list of employee &  & \multirow{2}{*}{\begin{tabular}[c]{@{}c@{}}Level of \\ inconvenience\end{tabular}} & restoring lost data &  \\ \cline{4-4} \cline{7-7}
 & \multicolumn{2}{c|}{} & financial data &  &  & perceive no inconvenience &  \\ \cline{2-4} \cline{6-7}
 & \multicolumn{2}{c|}{\multirow{2}{*}{Operational data}} & website availability &  & \multirow{3}{*}{Hinder operation} & financial lost &  \\ \cline{4-4} \cline{7-7}
 & \multicolumn{2}{c|}{} & operational system access &  &  & reputation damage &  \\ \cline{2-4} \cline{7-7}
 & \multicolumn{2}{c|}{\multirow{2}{*}{Intellectual property}} & \multirow{2}{*}{} &  &  & perceive no harm &  \\ \cline{6-7}
 & \multicolumn{2}{c|}{} &  &  & \begin{tabular}[c]{@{}c@{}}Attack experienced \\ by others\end{tabular} &  &  \\ \hline
\multirow{11}{*}{\textbf{Defense}} & \multicolumn{2}{c|}{\multirow{5}{*}{Technical}} & cloud/local backup & \multirow{11}{*}{\begin{tabular}[c]{@{}c@{}}\textbf{Information} \\ \textbf{Source}\end{tabular}} & \multirow{3}{*}{\begin{tabular}[c]{@{}c@{}}External \\ human\end{tabular}} & other institution &  \\ \cline{4-4} \cline{7-8} 
 & \multicolumn{2}{c|}{} & \multirow{2}{*}{\begin{tabular}[c]{@{}l@{}}version updates/\\ renew license\end{tabular}} &  &  & \multirow{2}{*}{IT consultant} & consulting agency \\ \cline{8-8} 
 & \multicolumn{2}{c|}{} &  &  &  &  & individual \\ \cline{4-4} \cline{6-8} 
 & \multicolumn{2}{c|}{} & security products &  & \multirow{5}{*}{\begin{tabular}[c]{@{}c@{}}External \\ non-human\end{tabular}} & \multirow{2}{*}{policy} & security standards \\ \cline{4-4} \cline{8-8} 
 & \multicolumn{2}{c|}{} & firewall &  &  &  & security regulations \\ \cline{2-4} \cline{7-8} 
 & \multicolumn{1}{c|}{\multirow{6}{*}{Organizational}} & \multirow{2}{*}{policy} & \multirow{2}{*}{\begin{tabular}[c]{@{}l@{}}security regulation/\\ security standard\end{tabular}} &  &  & conference \& lectures & \multirow{3}{*}{} \\ \cline{7-7}
 & \multicolumn{1}{c|}{} &  &  &  &  & journals &  \\ \cline{3-4} \cline{7-7}
 & \multicolumn{1}{c|}{} & \multirow{4}{*}{training} & incident response plan &  &  & news &  \\ \cline{4-4} \cline{6-8} 
 & \multicolumn{1}{c|}{} &  & employee training &  & \multirow{3}{*}{Personal} & education & \multirow{3}{*}{} \\ \cline{4-4} \cline{7-7}
 & \multicolumn{1}{c|}{} &  & phishing simulation &  &  & military service &  \\ \cline{4-4} \cline{7-7}
 & \multicolumn{1}{c|}{} &  & emergency drills &  &  & cyberattack experience &  \\ \hline
\end{tabular}%
}
\end{table}

\section{Awareness Level 1 and Level 4 Regression Coefficients}
\label{app:lv1_4_regress-coeff}

\begin{table}[h!]
\centering
\resizebox{\textwidth}{!}{%
\scriptsize
\begin{tabular}{lll|lll}
\specialrule{.2em}{.1em}{.1em}
\textbf{Level 1 Variables} & \textbf{Coefficients} & \textbf{Standard Errors} & \textbf{Level 4 Variables} & \textbf{Coefficients} & \textbf{Standard Errors} \\ \specialrule{.2em}{.1em}{.1em}
Has Cyber Insurance & \textbf{0.432*} & (0.194) & Has Cyber Insurance & \textbf{25.68***} & (15.75) \\
Revenue: 1-5 & 0.188 & (0.195) & Digital Assets: Customer list & 2.177 & (1.171) \\
Revenue: 5-10 & \textbf{0.120*} & (0.144) & Digital Assets: Customer financial data & 1.438 & (0.597) \\
Revenue: 10+ & \textbf{0.086**} & (0.105) & Digital Assets: Other customer sensitive data & \textbf{4.264***} & (2.048) \\
Revenue: undisclosed & \textbf{0.09***} & (0.084) & Digital Assets: Employee data & -0.577 & (0.282) \\
Sector: Professional Services & 0.516 & (0.399) & Digital Assets: Operational system & 1.135 & (0.609) \\
Sector: Trade & \textbf{4.976*} & (4.734) & Digital Assets: Operational data & 1.754 & (1.036) \\
Sector: Info. and Comm. & 1.064 & (0.869) & Digital Assets: Intellectual property & \textbf{2.724**} & (1.291) \\
Sector: Production & 1.677 & (1.579) & Digital Assets: Business financial data & 0.571 & (0.255) \\
\# of Digital Assets & 0.881 & (0.247) & Website: Hosting product or service information & 1.675 & (1.078) \\
Has a Website & 1.370 & (0.771) & Website: Selling product or service & 0.800 & (0.398) \\
Website: Hosting product or service information & 0.896 & (0.392) & Website: Displaying personalized content & 2.405 & (1.443) \\
Website: Selling product or service & \textbf{2.203**} & (0.765) & Website: Software as a service (SaaS) & 2.337 & (1.463) \\
Website: Displaying personalized content & 1.135 & (0.481) & Uses CRM or ERP & \textbf{5.443***} & (2.816) \\
Website: Software as a service (SaaS) & \textbf{2.360**} & (1.032) & Uses CRM or ERP: undisclosed & \textbf{4.701***} & (2.669) \\
Uses CRM or ERP & \textbf{2.250**} & (0.822) & Remote Work: Yes & 0.832 & (0.414) \\
Uses CRM or ERP: undisclosed & \textbf{2.527**} & (1.055) & Remote Work: No & \textbf{0.384*} & (0.221) \\
Revenue (1-5) X \# of Digital Assets & 1.374 & (0.432) & Program Installation: Cloud & 0.607 & (0.456) \\
Revenue (5-10) X \# of Digital Assets & \textbf{1.896*} & (0.680) & Program Installation: Cloud \& Local & \textbf{1.832*} & (0.837) \\
Revenue (10+) X \# of Digital Assets & 1.514 & (0.489) & Program Installation: Local & \textbf{0.310*} & (0.201) \\
Revenue (undisclosed) X \# of Digital Assets & \textbf{1.662*} & (0.47) & Program Installation: undisclosed & \textbf{0.117***} & (0.0772) \\
Sector (Prof. Service) X \# of Digital Assets & 1.280 & (0.261) & Constant & \textbf{215.9***} & (185.8) \\
Sector (Trade) X \# of Digital Assets & \textbf{0.489**} & (0.146) &  &  &  \\
Sector (Info. \& Comm.) X \# of Digital Assets & 1.151 & (0.242) &  &  &  \\
Sector (Production) X \# of Digital Assets & 0.774 & (0.166) &  &  &  \\
Constant & 0.607 & 0.574 &  &  & \\
\specialrule{.2em}{.1em}{.1em}
\end{tabular}%
}
    \begin{tablenotes}
        \footnotesize
        \item *** $p<0.01$, ** $p<0.05$, * $p<0.1$
    \end{tablenotes}
\end{table}

\newpage
\section{Business Characteristics Impacting Awareness of SMB Decision-Makers}
\label{app:awarenessprob}

\begin{table}[h!]
\centering
\resizebox{\textwidth}{!}{%
\begin{tabular}{l|ccccc|cc}
\specialrule{.2em}{.1em}{.1em}
\textbf{Variables} & \textbf{Low Awareness 1} & \textbf{Low Awareness 2} & \textbf{Low Awareness 3} & \textbf{Low Awareness 4} & \textbf{Low Awareness 5} & \textbf{2+ Low Awareness} & \textbf{No Low Awareness} \\ \specialrule{.2em}{.1em}{.1em}
Size: 6-10 & -0.0522 & \textbf{0.576*} & \textbf{1.023***} & 0.449 & 0.272 & 0.491 & 0.481 \\
 & (0.356) & (0.343) & (0.330) & (0.395) & (0.323) & (0.317) & (0.327) \\
Size: 51-100 & -0.491 & 0.256 & 0.402 & \textbf{1.150**} & \textbf{-0.775*} & -0.0893 & \textbf{0.729*} \\
 & (0.429) & (0.385) & (0.374) & (0.481) & (0.408) & (0.364) & (0.422) \\
Sector: Professional Services & -0.488 & -0.492 & -0.159 & 0.457 & \textbf{-0.651*} & \textbf{-0.857**} & 0.428 \\
 & (0.394) & (0.356) & (0.335) & (0.385) & (0.352) & (0.346) & (0.339) \\
Sector: Trade & -0.289 & 0.131 & -0.369 & -0.587 & -0.457 & -0.479 & 0.606 \\
 & (0.565) & (0.479) & (0.508) & (0.682) & (0.498) & (0.485) & (0.463) \\
Sector: Info. \& Comm. & 0.332 & \textbf{-1.075**} & \textbf{-1.045**} & 0.401 & 0.135 & -0.403 & 0.315 \\
 & (0.402) & (0.477) & (0.473) & (0.425) & (0.369) & (0.385) & (0.382) \\
Sector: Production & 0.123 & -0.0811 & 0.285 & 0.339 & \textbf{-1.026*} & 0.111 & 0.0508 \\
 & (0.513) & (0.446) & (0.419) & (0.509) & (0.545) & (0.416) & (0.447) \\
Revenue: 1-5 & -0.155 & 0.0228 & -0.231 & -0.183 & -0.731 & -0.765 & 0.153 \\
 & (0.516) & (0.537) & (0.527) & (0.543) & (0.512) & (0.486) & (0.535) \\
Revenue: 5-10 & -0.766 & -0.484 & -0.261 & -0.792 & -0.384 & \textbf{-1.103**} & 0.747 \\
 & (0.598) & (0.643) & (0.599) & (0.625) & (0.549) & (0.561) & (0.560) \\
Revenue: 10+ & -0.637 & 0.268 & 0.454 & -0.822 & 0.176 & -0.318 & -0.313 \\
 & (0.567) & (0.563) & (0.558) & (0.583) & (0.518) & (0.505) & (0.579) \\
Revenue: Undisclosed & -0.765 & 0.0436 & -0.269 & \textbf{-0.885*} & -0.411 & \textbf{-0.903**} & 0.287 \\
 & (0.488) & (0.493) & (0.487) & (0.521) & (0.450) & (0.442) & (0.492) \\
Tech. Intensity: High & 0.532 & -0.0366 & \textbf{-0.868**} & 0.250 & 0.481 & -0.495 & \textbf{-0.524*} \\
 & (0.324) & (0.330) & (0.348) & (0.331) & (0.296) & (0.316) & (0.309) \\
Experienced Cyberattack & -0.166 & \textbf{-1.190***} & \textbf{-0.720**} & \textbf{-0.618*} & 0.0456 & \textbf{-0.967***} & \textbf{0.617**} \\
 & (0.348) & (0.382) & (0.343) & (0.374) & (0.312) & (0.343) & (0.287) \\
Constant & \textbf{-0.920*} & -0.783 & -0.665 & \textbf{-1.405**} & -0.636 & -1.780** & 0.362 \\
 & (0.515) & (0.525) & (0.518) & (0.551) & (0.482) & (0.517) & (0.473) \\ \hline
Total Percentage & 20\% & 23\% & 27\% & 20\% & 25\% & 34\% & 29\% \\
\specialrule{.2em}{.1em}{.1em}
\end{tabular}%
}
    \begin{tablenotes}
        \footnotesize
        \item Standard errors in parentheses
        \item *** $p<0.01$, ** $p<0.05$, * $p<0.1$
    \end{tablenotes}
\end{table}

\end{document}